\documentclass[12pt]{article}
 \usepackage{fullpage}

\usepackage[pdftex]{graphicx}
\usepackage{amsfonts}
\usepackage{amscd}
\usepackage{indentfirst}
\usepackage{amssymb,amsthm, amsgen, amstext, amsbsy, amsopn}
\IfFileExists{yfonts.sty}{\usepackage{yfonts}}{}
\usepackage{xcolor}

 \usepackage{cancel}
\usepackage[normalem]{ulem}
 
\newif\ifshowcorrections
\showcorrectionstrue
\newsavebox{\correctionbox}

\usepackage{amsmath}
\usepackage{array}
\usepackage{mathrsfs}

\usepackage{booktabs}
\usepackage[margin=1in]{geometry}

\usepackage[hyphens]{url}
\usepackage[colorlinks=true,pdfpagemode=UseNone,urlcolor=blue,linkcolor=blue,citecolor=blue,breaklinks=true]{hyperref}
\usepackage{cleveref}
\newtheorem{thm}{Theorem}[section]
\newtheorem{lemma}[thm]{Lemma}
\newtheorem{prop}[thm]{Proposition}

\newtheorem{rem}[thm]{Remark}

\newcommand{\N}{{\mathbb{N}}}

\newcommand{\R}{{\mathbb{R}}}

\newcommand{\cN}{{\mathcal{N}}}

\newcommand{\Qed}{\ \hfill \qedsymbol \bigskip}

\newcommand{\bal}{\begin{aligned}}
\newcommand{\enbal}{\end{aligned}}
\newcommand{\be}{\begin{equation}}
\newcommand{\ee}{\end{equation}}
\newcommand{\farc}{\frac}
\newcommand{\pdr}[2]{\frac{\partial{#1}}{\partial{#2}}}

\newcommand{\Rm}{{\mathbb R}}
\newcommand{\Zm}{{\mathbb Z}}

\newcommand{\Em}{{\mathbb E}}
\newcommand{\Nm}{{\mathbb N}}
\newcommand{\Pm}{{\mathbb P}}
\newcommand{\eps}{\varepsilon}

\numberwithin{equation}{section}
\newcommand{\msfw}{{\mathsf w}}

\begin{document}

\title{Characterizing Ultrafast Processes via Work}
\author{Bernard Derrida$^{1,2}$, Michael Entov$^3$,  Leonid Polterovich$^4$ and Lenya Ryzhik$^5$}

\maketitle

\footnotetext[1] {Coll\`ege de France, France}
\footnotetext[2] {Laboratoire de Physique de l'\'Ecole Normale Sup\'erieure, ENS,
Universit\'e PSL, CNRS, Sorbonne Universit\'e, Universit\'e de Paris, France}
\footnotetext[3]{Faculty of Mathematics, Technion -- Israel Institute of
Technology, Israel}
\footnotetext[4]{School of Mathematical Sciences, Tel Aviv University, Israel}
\footnotetext[5]{Department of Mathematics, Stanford University, USA.}

%%%%%%%%%%%%%
%%%%%%
\begin{abstract}
Consider the following cycle for a mean-field system initially in thermodynamic equilibrium: an instantaneous parameter jump, relaxation to the new equilibrium, and a reverse jump. For certain systems and parameter values, the two equilibrium macrostates coincide in the thermodynamic limit.

We characterize this matching phenomenon through the work required to perform the cycle. Our main result concerns scalar many-particle mean-field systems under suitable assumptions. In the matching regime, the total work has a non-degenerate, non-Gaussian limiting distribution, and the average work per particle tends to zero. Away from matching, the properly rescaled fluctuations of the total work have a possibly degenerate Gaussian limiting distribution, and the average work per particle converges to a strictly positive constant. Further motivation arises from a contact-geometric model of thermodynamics, in which matching corresponds to a Reeb chord between two equilibrium Legendrian submanifolds.
\end{abstract}

\tableofcontents

\section{Introduction}

In many situations in physics, it is possible to change the Hamiltonian
of a system almost instantaneously. This is, for example,
the case when a field acting on a magnetic system
is switched off. Another case would be a fast change of the confining potential on atoms or
molecules using optical tweezers~\cite{BCLW,tweezers-survey,Grier}. During such instantaneous changes,
the system has no time to
exchange heat with heat baths, so that the entire change in the system's energy is due to
work performed on it. In this paper, we consider the evolution of a system subjected to
a sequence of such changes of its Hamiltonian, which between any two consecutive
jumps has enough time to relax to the new equilibrium. In particular, we want to determine
the work performed on the system over a cycle consisting of a jump from one
Hamiltonian to another, relaxation to the new equilibrium, and a reverse
jump followed by relaxation to the original equilibrium.

We will be especially interested in instantaneous changes between two mean field
Hamiltonians in the regimes where the Gibbs laws of the macroscopic
variable converge weakly, in the thermodynamic limit, to the same delta
measure, despite the fact that the two Hamiltonians are different.
We call such a transition an {\it ultrafast process} in the macroscopic
sense: the instantaneous change of Hamiltonian leaves the limiting
equilibrium value of the macroscopic variable unchanged. This terminology
refers to equilibrium matching in the thermodynamic limit and makes no
assertion about the time required for the finite system to equilibrate.
Ultrafast processes have previously been related to Reeb chords in
symplectic topology in~\cite{EPR-JSP,EPR-LMP}. In the contact-geometric
model of thermodynamics considered there, these chords connect the
equilibrium Legendrian submanifolds of the two systems.

Our main result concerns scalar $N$-particle mean-field systems.
We show that, under the assumptions below,
another signature of an ultrafast process is the emergence of a
non-degenerate, non-Gaussian limiting distribution of the total work
performed on the system over the cycle. The limiting random variable
is a quadratic expression in two independent Gaussian variables.
The corresponding average work per particle is of order $N^{-1}$,
with an explicit strictly positive coefficient, where $N$ is the
number of particles in the mean-field system. On the other hand,
away from matching, the fluctuations of the total work about its
leading deterministic term have a possibly degenerate Gaussian limit
after rescaling by~$N^{-1/2}$, and the average work per particle
converges to a strictly positive constant.
We also express the average work as a symmetrized relative entropy
and compute the asymptotics of the standard deviation of the work
per particle.

%%%%%%%%%%%%%

 \subsection{The jump--relaxation--jump cycle and positivity of work}
\label{sec:cycle}

Let $E_0,E_1:X\to\mathbb R$
be two energy functions defined on a
probability  measure space~$(X,\rho)$,
and~$\beta>0$ be a parameter (in physical models it stands for the inverse of
the temperature),
such that the corresponding partition functions
\[
 Z_i=\int_X e^{-\beta E_i(x)}\,d\rho(x)
\]
are finite.  The Gibbs measures corresponding to the two energies $E_{0,1}$ are
\[
 d\mu_i(x)=\frac1{Z_i}e^{-\beta E_i(x)}\,d\rho(x),~~x\in X,~~i=0,1.
\]
We assume that the difference $E_1-E_0$ is integrable with respect to
both Gibbs measures. We model the microscopic states of the system immediately before the first and second jumps by independent random variables with laws \(\mu_0\) and \(\mu_1\), respectively.

Consider the following cycle.  The system starts with the equilibrium
Gibbs measure~$\mu_0$ corresponding to the energy function $E_0(x)$.  The energy changes instantaneously from
$E_0(x)$ to $E_1(x)$, while the microscopic state $x\in X$ is still distributed according to the measure~$\mu_0$.  The system then relaxes
under the energy $E_1(x)$ to the Gibbs measure~$\mu_1$.  Next, the energy
changes instantaneously back from $E_1$ to $E_0$, while the microscopic state is
distributed according to the measure $\mu_1$.
After the second jump, the system relaxes back to the Gibbs measure $\mu_0$.

The two relaxations perform no work, and the change of energy of a microstate $x\in X$
at the jump is
\be
\Delta E(x):=E_1(x)-E_0(x),
\ee
when one switches from $E_0(x)$ to $E_1(x)$, and $(-\Delta E(x))$ when one switches back.
At the microscopic level, the work over the cycle is, therefore,
\be\label{26oct2508}
w(x,x')=\Delta E(x)+(-\Delta E(x'))=E_1(x)-E_0(x)+E_0(x')-E_1(x').
\ee
Here the first term corresponds to the jump in energy at the original microscopic state $x\in X$ and the second to the jump of energy at the second microscopic state $x'\in X$
that resulted  from the first relaxation process.
Hence, the total average work performed
on the system over the two jumps is
\begin{equation} \label{26oct3002}
\bal
 W &=\int_X(E_1(x)-E_0(x))\,d\mu_0(x)
 -\int_X(E_1(x')-E_0(x'))\,d\mu_1(x')\\
 &=\int_{X\times X}w(x,x')d\mu_0(x)d\mu_1(x').
\enbal
\end{equation}

We now explain why the average work over such cycle is always
non-negative,
as it should be according to the Second Law.
Recall that, for probability measures $\mu$ and $\nu$ defined on the same measure space $X$,
with $\mu$ being absolutely continuous with respect to $\nu$,
their relative entropy is
\[
 D(\mu\|\nu)
 =
 \int_X
 \log\Big(\frac{d\mu}{d\nu}\Big)\,d\mu.
\]
It satisfies $D(\mu\|\nu)\geq0$, with equality if and only if $\mu=\nu$, by Jensen's inequality.

\begin{prop}[Positivity of work]
\label{prop:work-positive}
The total average work over the cycle is given by
\begin{equation} \label{26oct2910}
 W=
 \farc{1}{\beta}\big[D(\mu_0\|\mu_1)+D(\mu_1\|\mu_0)\big].
 \end{equation}
Thus, the total average work over the cycle is always non-negative: $W\ge 0$, and $W=0$ if and only if $\mu_0=\mu_1$.
%\[
% W\geq0,
% \qquad
% W=0
% \quad\Longleftrightarrow\quad
% \mu_0=\mu_1.
%\]
%In particular, the work is strictly positive whenever the two Gibbs
%measures are different.
\end{prop}

\begin{proof}
Since
\[
 \log\frac{d\mu_0}{d\mu_1}(x)
 =
 \beta(E_1(x)-E_0(x))+\log\frac{Z_1}{Z_0},
\]
and $\mu_0$ is a probability measure, we have
\[
 D(\mu_0\|\mu_1)
 =
 \beta\int_X(E_1(x)-E_0(x))\,d\mu_0
 +
 \log\frac{Z_1}{Z_0}.
\]
Similarly, we have
\[
 D(\mu_1\|\mu_0)
 =
 \beta\int_X(E_0(x)-E_1(x))\,d\mu_1
 +
 \log\frac{Z_0}{Z_1}.
\]
Adding the two identities gives \eqref{26oct2910}.
\end{proof}

We now specialize to the scalar mean-field systems studied in this paper.
Let $s_1,\ldots,s_N$ be independent identically distributed real-valued
random variables with a common continuous probability density $\rho(s)$,
and let $P_N$ be the law of their empirical mean
\begin{equation}\label{26oct2812}
 m=\frac1N\sum_{j=1}^N s_j.
\end{equation}
For $i=0,1$, let $H_i(m)$ be the energy per particle, so that the
$N$-particle Hamiltonian is~$NH_i(m)$. The equilibrium law of the
empirical mean is the Gibbs measure
\begin{equation}\label{26oct2914}
 d\nu_i^{(N)}(m)
 =\frac1{Z_{i,N}}e^{-\beta N H_i(m)}\,P_N(dm),
 \qquad i=0,1,
\end{equation}
where
\begin{equation}\label{26sep1402}
 Z_{i,N}=\int_{\mathbb R}e^{-\beta N H_i(m)}\,P_N(dm).
\end{equation}
Let $M_0^{(N)}$ and
$M_1^{(N)}$ be independent random variables with laws $\nu_0^{(N)}$
and $\nu_1^{(N)}$, respectively.

The work per particle {over the cycle}, evaluated at the two macroscopic states is
\begin{equation}\label{26oct2318}
 w(m,m')=\Delta H(m)-\Delta H(m'),
\end{equation}
where
\begin{equation}\label{eq:energy-difference}
 \Delta H(m)=H_1(m)-H_0(m),\qquad m\in\mathbb R,
\end{equation}
and the random work per particle over the cycle is
\begin{equation}\label{eq:random-work}
 \mathsf w_N=w(M_0^{(N)},M_1^{(N)}).
\end{equation}
Its expectation {and variance are}
\begin{equation}\label{26oct2934}
 \begin{aligned}
 w_N:=\mathbb E[\mathsf w_N]
 &=\int_{\mathbb R}\Delta H(m)\,d\nu_0^{(N)}(m)
   -\int_{\mathbb R}\Delta H(m)\,d\nu_1^{(N)}(m)\\
 &=\int_{\mathbb R^2}w(m,m')\,
       d\nu_0^{(N)}(m)\,d\nu_1^{(N)}(m'),
 \end{aligned}
\end{equation}
{and}
\begin{equation}\label{eq:work-standard-deviation}
 \sigma_N:=\sqrt{\operatorname{Var}(\mathsf w_N)}
 =\left(\mathbb E[(\mathsf w_N-w_N)^2]\right)^{1/2}.
\end{equation}
The assumptions in Section~\ref{sec:scalar-case} below ensure that
the partition functions are finite and positive and that the moments
used here are finite for all sufficiently large $N$.
Proposition~\ref{prop:work-positive}, applied with $E_i=NH_i$, gives
$w_N\ge0$.

Our aim is to determine the limiting distribution of the suitably
rescaled work, the asymptotics of its expectation, and the size of its
fluctuations as $N\to\infty$. We will see a sharp distinction between
the cases where the two limiting equilibrium macrostates differ and
where they coincide.

 \subsection{Work cycle in mean-field systems}
\label{sec:scalar-case}

We now state the assumptions on the reference density $\rho$ and the
energies per particle~$H_0,H_1$ in the probabilistic setting of
Section~\ref{sec:cycle}.

\subsubsection{Two remarkable functions}

We assume that {\it the log-moment generating function}
\begin{equation}\label{26oct2932}
 \Lambda(\theta)=\log\int_{\mathbb R}e^{\theta s}\rho(s)\,ds
\end{equation}
is finite in a neighborhood of~$\theta=0$. We also assume, for convenience,
that its domain
\begin{equation}\label{26sep1408}
 D=\{\theta\in\mathbb R:\Lambda(\theta)<+\infty\}
\end{equation}
is an open interval  $(-a,b)$, where $a,b\in(0,+\infty]$.
\color{black} This assumption ensures that the rate function $I$ defined below in
\eqref{26oct2306} is smooth and strictly convex wherever it is finite, with the supremum in its definition attained at an interior point of \(D\). \color{black}

{\it The Cram\'er rate function} is the Legendre transform of~$\Lambda$:
\begin{equation}\label{26oct2306}
 I(m)=\sup_{\theta\in\mathbb R}\{\theta m-\Lambda(\theta)\}.
\end{equation}
\color{black} It governs the large deviations of the empirical mean
$m=N^{-1}\sum_{i=1}^N s_i$. We refer to  Section~\ref{sec:interpret} for further details.\color{black}

Set
\begin{equation}\label{26oct3028}
 \bar s_-=\inf\{s\in\mathbb R:\rho(s)>0\},\qquad
 \bar s_+=\sup\{s\in\mathbb R:\rho(s)>0\},
\end{equation}
and write
\begin{equation}\label{26oct3027}
 D'=(\bar s_-,\bar s_+).
\end{equation}
Either endpoint of~$D'$ may be infinite.
The following proposition collects the properties used below.
Its proof is given in Appendix~\ref{app:log-mgf-rate}.

\begin{prop}\label{prop:log-mgf-rate}
Under the assumptions above, the following statements hold.
\begin{enumerate}
\item[\rm (i)]
The function~$\Lambda$ is $C^\infty$ on~$D$,
\begin{equation}\label{26oct2306-1}
 \Lambda''(\theta)>0,\qquad \theta\in D,
\end{equation}
and
\begin{equation}\label{26oct3029}
 D'=\{\Lambda'(\theta):\theta\in D\}.
\end{equation}

\item[\rm (ii)]
The function~$I$ is smooth  on~$D'$, and
\begin{equation}\label{26oct3026}
 I(m)=+\infty,\qquad m\notin D'.
\end{equation}

\item[\rm (iii)]
For every $m\in D'$, the supremum in~\eqref{26oct2306} is attained at
a unique point $\bar\theta(m)\in D$, characterized by
\begin{equation}\label{26oct2502}
 m=\Lambda'(\bar\theta(m)).
\end{equation}
Moreover, for all $m\in D'$,
\be\label{26oct2310}
\bal
 &I(m)=m\bar\theta(m)-\Lambda(\bar\theta(m)),\\
 &I'(m)=\bar\theta(m).
 \enbal
\end{equation}
Furthermore, $I$ is strictly convex on $D'$:
\begin{equation}\label{26oct2629}
 I''(m)=\bar\theta'(m)
 =\frac{1}{\Lambda''(\bar\theta(m))}>0.
\end{equation}
\end{enumerate}
\end{prop}

\subsubsection{Assumptions on the energy per particle}

We assume throughout that both energies per particle $H_0,H_1$ are
of the smoothness class~$C^2$. As in Section~\ref{sec:cycle}, they depend on the empirical mean
\eqref{26oct2812}, and the corresponding $N$-particle Hamiltonians
are $NH_i(m)$, $i=0,1$. We impose a number of assumptions to apply the Laplace method in the setting of large deviations. This method is the main technical tool used in the paper.

\begin{itemize}

\item{}
$H_0$ and $H_1$  are bounded from below: there
exist $D_i>0$ so that
\be\label{26oct2320}
H_i(m)\ge -D_i,~~\hbox{for all $m\in\Rm$ and $i=0,1$,}
\ee

\item{}
$H_0$ and $H_1$
 have at most exponential
growth: there exist constants $C_i,\kappa_i>0$ such that
\be\label{26oct2602}
 |H_i(m)|\leq C_i e^{\kappa_i|m|},
 \qquad m\in\mathbb R.
\ee

\end{itemize}

We also define the function
\be\label{26oct2302}
 \Phi_i(m)=I(m)+\beta H_i(m),
 \ee
with some inverse temperature $\beta>0$. Its meaning is also discussed in
Section~\ref{sec:interpret} below. We assume that $\Phi_i$, $i=0,1$ have
the following property.

\bigskip
\noindent
(A) For each $i=0,1$, the function $\Phi_i(m)$ has a unique global
minimizer $m_i$. This minimizer is non-degenerate,
\begin{equation}\label{26oct2912}
A_i:=\Phi_i''(m_i)>0,
%=\frac{1}{\sigma_i^2}+\beta H_i''(m_i)>0.
\end{equation}
and
\be\label{26oct2323}
A_0\neq A_1.
\ee
In principle, our results below hold also for $A_0=A_1$. This assumption
is made solely so that the leading order terms in some computations below
do not vanish.

An important example of a pair of systems satisfying these assumptions, {\it Gaussian oscillators},
is elaborated in Section \ref{sec:gaussian-example}.

As the minimum value of $\Phi_i (m)$, $i=0,1$, reached at $m=m_i$, is finite,
we know from~\eqref{26oct3026} that $m_i\in D'$.
We denote
\be\label{26oct2316}
\theta_i=\bar\theta(m_i),
\ee
so that
\be\label{26oct2314}
m_i=\Lambda'(\theta_i),
\ee
because of~\eqref{26oct2502}.
As $m_i$ is a critical point of $\Phi_i(m)$ and because of~\eqref{26oct2310} and~\eqref{26oct2316},
we also have
\be\label{26oct2635}
\theta_i=-\beta H_i'(m_i).
\ee

 \subsubsection{An interpretation of $I(m)$ and $\Phi_i(m)$}
\label{sec:interpret}

Recall that $P_N$ is the law of the empirical mean under the product
reference measure, while $\nu_i^{(N)}$ is its Gibbs law
\eqref{26oct2914} for the Hamiltonian $NH_i$.
The points $m_0$ and $m_1$ are the limiting equilibrium macroscopic states: we will see
in Lemma~\ref{lem:26oct2704} below that
the measures~$\nu_i^{(N)}$ concentrate at $m=m_i$, as~$N\to +\infty$.
Thus, the condition $m_0=m_1$ is precisely the matching condition relevant to an ultrafast
process {\sl in the thermodynamic limit}.

Let us now explain the meaning of the functions $I(m)$ and $\Phi_i(m)$. In terms of large deviations,
the function $I(m)$ is the  rate function for the measures $P_N(dm)$, in the sense that
for every closed set
$F\subset \Rm$ and every open set $G\subset \Rm$, we have
\[
 \limsup_{N\to +\infty}\frac1N\log P_N(F)
 \leq-\inf_F I,
\]
and
\[
 \liminf_{N\to +\infty}\frac1N\log P_N(G)
 \geq-\inf_G I.
\]
This is the claim of Cram\'er's theorem, see Theorem~\ref{thm:cramer} in Appendix~\ref{app:large-dev}.
In a similar vein, a consequence of Varadhan's lemma, Theorem~\ref{thm-lem-varadhan} in Appendix~\ref{app:large-dev}, is that
\be
J_i(m):=\Phi_i(m)-\Phi_i(m_i)
\ee
is the  rate function for the Gibbs measures $\nu_i^{(N)}(dm)$.

Thus, up to a constant, $\Phi_i(m)$ is the rate function for the Gibbs measure $\nu_i^{(N)}(dm)$.
This leads to a more physical interpretation of $\Phi_i(m)$. For $i=0,1$, let
\[
f_{i,N}:=-\frac{1}{\beta N}\log Z_{i,N}
\]
be the equilibrium free energy per particle corresponding to the Hamiltonian $NH_i$.
Once again, by Varadhan's lemma,  and the uniqueness of the minimizer $m_i$ of $\Phi_i$, we have
\be\label{26oct2325}
 f_i:=\lim_{N\to +\infty}f_{i,N}
 =H_i(m_i)+\frac1\beta I(m_i)=\frac1\beta\Phi_i(m_i).
\ee
Therefore, $\Phi_i(m_i)$ is a multiple of the limiting (as $N\to+\infty$) free energy per particle.

\subsection{Main theorem}\label{sec:work}

We use the independent random variables $M_0^{(N)},M_1^{(N)}$ and the
random work $\mathsf w_N$, its expectation $w_N$, and its standard
deviation $\sigma_N$ introduced in Section~\ref{sec:cycle}.
The limiting equilibrium macrostates $m_i$ and the constants $A_i>0$
are specified in assumption~(A). Recall that
\begin{equation}\label{26oct2320bis}
 \Delta H(m)=\frac1\beta[\Phi_1(m)-\Phi_0(m)],\qquad m\in D'.
\end{equation}
Set
\begin{equation}\label{26oct2728}
 d_i=\Delta H'(m_i),\qquad i=0,1,
\end{equation}
and write
\be\label{26nov1902}
 w_\infty=\Delta H(m_0)-\Delta H(m_1),
 \qquad V=\frac{d_0^2}{A_0}+\frac{d_1^2}{A_1}.
\ee
In the matching case $m_0=m_1=m_*$, we also set
\be\label{26nov1904}
 c=\Delta H''(m_*)=\frac{A_1-A_0}{\beta}.
\ee

We write $X_N\Longrightarrow X$ for convergence in distribution,
that is, weak convergence of the probability laws of $X_N$ to
the law of $X$.

\begin{thm}[Distribution, expectation, and fluctuations of work]
\label{thm:main}
Under the assumptions of Sections~\ref{sec:cycle} and~\ref{sec:scalar-case},
the following statements hold as $N\to\infty$.

\begingroup
% The former theorem labels now refer to the corresponding items.
\renewcommand{\theenumi}{\thethm(\roman{enumi})}
\renewcommand{\labelenumi}{\textup{(\roman{enumi})}}
\begin{enumerate}
\item\label{thm:work-distribution}
\emph{Convergence in distribution.}
If $m_0\ne m_1$, then
\begin{equation}\label{eq:work-distribution-nonmatching}
 \sqrt N\,(\mathsf w_N-w_\infty)
 \Longrightarrow\mathcal N(0,V),
\end{equation}
with $V$ defined in~\eqref{26nov1902}.
The limiting Gaussian is non-degenerate\footnote{Here a probability law
is called non-degenerate if it is not concentrated at a single point,
or equivalently, if the corresponding random variable is not almost
surely constant.} when at least one of $d_0,d_1$ defined in~\eqref{26oct2728} is nonzero.
If $m_0=m_1=m_*$, then
\begin{equation}\label{eq:work-distribution-matching}
 N\mathsf w_N\Longrightarrow
 \zeta:=\frac{c}{2A_0}\xi_0^2-\frac{c}{2A_1}\xi_1^2,
\end{equation}
where $\xi_0,\xi_1$ are independent standard normal random variables,
the numbers $A_{0,1}$ are defined in~\eqref{26oct2728}, and $c$ in~\eqref{26nov1904}.
Under our standing assumption~\eqref{26oct2323}, $c\neq 0$, the law of $\zeta$ is
non-Gaussian and non-degenerate.

\item\label{thm:work-asympt}
\emph{Convergence of expectations.}
If $m_0\ne m_1$, then
\begin{equation}\label{26oct2716}
 w_N=w_\infty+o(1),\qquad w_\infty>0.
\end{equation}
If $m_0=m_1=m_*$, then
\begin{equation}\label{26oct2718}
 \begin{aligned}
 w_N
 &=\frac{c}{2N}\left(\frac1{A_0}-\frac1{A_1}\right)+o(N^{-1})
 =\frac{(A_1-A_0)^2}{2\beta N A_0A_1}+o(N^{-1}).
 \end{aligned}
\end{equation}
Thus, the average work per particle is of order one when the
equilibrium macrostates differ and of order $N^{-1}$ at matching.

\item\label{thm:work-fluct}
\emph{Convergence of standard deviations.}
If $m_0\ne m_1$, then
\begin{equation}\label{26oct2730}
 \sqrt N\,\sigma_N\longrightarrow\sqrt V
 =\left(\frac{d_0^2}{A_0}+\frac{d_1^2}{A_1}\right)^{1/2}.
\end{equation}
If $m_0=m_1=m_*$, then
\begin{equation}\label{26oct2731}
 N\sigma_N\longrightarrow\sqrt{\operatorname{Var}(\zeta)}
 =\frac{|c|}{\sqrt2}
   \left(\frac1{A_0^2}+\frac1{A_1^2}\right)^{1/2}.
\end{equation}
\end{enumerate}
\endgroup
\end{thm}

The distributional limits in part~\textup{(i)} also hold with convergence
of all moments (see Remark \ref{rem-allmoments} below). In particular, the limits of expectations and standard deviations are consistent with the limiting laws.

\begin{rem}\label{rem:zero-variance}{\rm
It is possible that $V=0$ even when $m_0\ne m_1$.
This happens when $\Phi'_1(m_0) = \Phi'_0(m_1) = 0$.
In this case, the law of
\[
 \sqrt N\,(\mathsf w_N-w_\infty)
\]
converges weakly to the Dirac measure at zero, or equivalently,
this random variable converges to zero in probability.
Thus the total work $N\mathsf w_N$ has a positive leading term
$Nw_\infty$, but its fluctuations are smaller than $\sqrt N$.
Moreover, part~\textup{(iii)} of the theorem gives
\[
 N\sigma_N=o(\sqrt N)
\]
for the standard deviation of the total work.
The theorem does not determine the finer scale of these fluctuations
or their first non-degenerate limiting distribution.
}
\end{rem}

\begin{rem}[Jarzynski's identity]\label{rem:jarzynski}{\rm
The characteristic function of the matching limit is
\begin{equation}\label{eq:work-distribution-characteristic}
 \mathbb E[e^{it\zeta}]
 =\left(1-\frac{ict}{A_0}\right)^{-1/2}
  \left(1+\frac{ict}{A_1}\right)^{-1/2},\qquad t\in\mathbb R,
\end{equation}
where the square roots are chosen continuously with value $1$ at $t=0$.
The same Gaussian integral is finite at $t=i\beta$, since
$c=(A_1-A_0)/\beta$ gives
\[
 1+\frac{\beta c}{A_0}=\frac{A_1}{A_0}>0,
 \qquad
 1-\frac{\beta c}{A_1}=\frac{A_0}{A_1}>0.
\]
Thus, taking $it=-\beta$ in~\eqref{eq:work-distribution-characteristic},
we obtain
\[
 \mathbb E[e^{-\beta\zeta}]
 =\left(\frac{A_1}{A_0}\right)^{-1/2}
  \left(\frac{A_0}{A_1}\right)^{-1/2}=1.
\]
Hence the limiting total work satisfies the cyclic form of
Jarzynski's identity~\cite{Jar97}, in which the free energy difference
between the initial and final equilibrium states is zero.
}
\end{rem}

When $m_0\ne m_1$ and $V>0$, the total work $N\mathsf w_N$ has a
positive leading term $Nw_\infty$ and asymptotically Gaussian
fluctuations of order $\sqrt N$. At matching, the total work instead
converges to the non-Gaussian random variable $\zeta$. Thus equilibrium
matching does not eliminate fluctuations of the total work: the two
equilibrium measures have different fluctuation variances, producing
random work of order one even though the work per particle tends to zero.

Since $w_\infty=0$ in the matching case, both distributional limits
concern $\mathsf w_N-w_\infty$, the work relative to its limiting mean,
rescaled by $\sqrt N$ when $m_0\ne m_1$ and by $N$ when $m_0=m_1$.
The following table summarizes the theorem; its last row refers to
the limiting distributions after these rescalings.
\begin{center}
\renewcommand{\arraystretch}{1.25}
\begin{tabular}{@{}lcc@{}}
\toprule
 & $m_0\ne m_1$ & $m_0=m_1=m_*$\\
\midrule
Average work per particle & $O(1)$ & $O(N^{-1})$\\
Standard deviation & $O(N^{-1/2})$ & $O(N^{-1})$\\
Limiting distribution & Gaussian & Non-Gaussian\\
\bottomrule
\end{tabular}
\end{center}
The standard-deviation order in the nonmatching case is attained when
$V>0$; if $V=0$, the Gaussian limit is a point mass and
$\sigma_N=o(N^{-1/2})$. The matching orders are attained by
$A_0\ne A_1$.
The \(N^{-1}\) fluctuation scale at matching reflects the identity
$\Delta H'(m_*)=0$: a deviation of the empirical mean of size
$N^{-1/2}$ contributes quadratically to the jump work, hence at
order $N^{-1}$.

For completeness, the positivity of $w_\infty$ in the nonmatching
case follows directly from
\[
 w_\infty=\frac1\beta\bigl[\Phi_0(m_1)-\Phi_0(m_0)
                    +\Phi_1(m_0)-\Phi_1(m_1)\bigr]>0,
\]
because $m_i$ is the unique minimizer of $\Phi_i$.
At matching, the coefficient in~\eqref{26oct2718} is positive by
$A_0\ne A_1$. If this last assumption is dropped and $A_0=A_1$,
then $\zeta=0$, $w_N=o(N^{-1})$, and~$\sigma_N=o(N^{-1})$;
a finer asymptotic analysis is needed to determine the next scale.

\begin{rem}\label{rem-free}{\rm
In the matching case $m_0=m_1=m_*$, we see from~\eqref{26oct2325} that
\[
 f_1-f_0=H_1(m_*)-H_0(m_*)=\Delta H(m_*).
\]
Thus the jump in the energy per particle is precisely the difference
of the limiting equilibrium free energies per particle.
}
\end{rem}

\subsection{Further generalizations}\label{sec:generalizations}

The scalar result raises the question of what happens when the energy
depends on several macroscopic variables. We briefly outline how the asymptotics of the expected work extend to this setting.  Consider the observables
\[
 \vec v(s)=(s,s^2,\ldots,s^K),\qquad s\in\mathbb R,
\]
and the macroscopic state
\[
 \vec m=\frac1N\sum_{j=1}^N\vec v(s_j)\in\mathbb R^K.
\]
Let the two Hamiltonians be $NH_i(\vec m)$, and let $I(\vec m)$ be
the Cram\'er rate function for this empirical mean. Assume the
multidimensional analogues of the hypotheses above, including the
necessary exponential-moment assumptions on $\vec v(s)$, and suppose
that
\[
 \Phi_i(\vec m)=I(\vec m)+\beta H_i(\vec m),\qquad i=0,1,
\]
have unique non-degenerate global minimizers $\vec m_i$. Set
\[
 A_i=\nabla^2 I(\vec m_i)+\beta\nabla^2 H_i(\vec m_i),
 \qquad \Delta H=H_1-H_0.
\]
Here $A_0,A_1$ are positive definite matrices. As in the scalar case,
we take the macroscopic states at the two jumps to be independent,
with their respective Gibbs laws, and write $w_N$ for the expected
work per particle.

For expectations, we have the same dichotomy as in
Theorem~\ref{thm:work-asympt}:
\begin{equation}\label{26oct2806}
 w_N=\Delta H(\vec m_0)-\Delta H(\vec m_1)+o(1).
\end{equation}
If $\vec m_0\ne\vec m_1$, the limiting constant is strictly positive.
If $\vec m_0=\vec m_1=\vec m_*$, the analogous second-order
calculation gives
\begin{equation}\label{26oct2808}
 w_N=\frac1{2\beta N}
       \operatorname{Tr}(B+B^{-1}-2\mathbb I)+o(N^{-1}),
 \qquad B=A_0^{-1/2}A_1A_0^{-1/2}.
\end{equation}
The leading coefficient is non-negative and vanishes precisely
when $A_0=A_1$.

This result for expectations suggests the following question:
what are the corresponding limiting distributions and fluctuation
asymptotics? In particular, does the Gaussian versus non-Gaussian
dichotomy persist, and which quadratic forms describe the limiting
work at matching? We leave these questions for further study.

 \subsection{Connection to Reeb chords} \label{sec:reeb}

%\begin{rem}\label{rem-param} {\rm
Assume that the functions $H_i(m)$, $i=0,1$, depend on an external parameter
$q\in\mathbb R$ as follows:
\begin{equation}\label{eq-param}
H_i(m,q)=h_i(m)+mq.
\end{equation}
For each $q$ and $i=0,1$, denote by $m_i(q)$ the minimizer of
\[
\Phi_i(m,q)=I(m)+\beta H_i(m,q).
\]
We assume that our previous assumptions on the energy per particle hold for
$H_i(m,q)$, $i=0,1$, for all $q$, and that the functions  $m_i(q)$, $i=0,1$,
are differentiable. We set
\be\label{26oct2510}
f_i(q)
:=
\frac1\beta\Phi_i(m_i(q),q)
=
\min_{m\in\Rm}\left\{
\frac1\beta I(m)+h_i(m)+mq
\right\}.
\ee
By \eqref{26oct2325}, $f_i(q)$, $i=0,1$, is the limiting, as $N\to +\infty$,
equilibrium free energy per particle of the $i$-th system at the parameter value $q$.

Since $m_i(q)$ is a critical point of $\Phi_i(m,q)$,
differentiation of~\eqref{26oct2510} with respect to $q$ gives
\[
f_i'(q)=\farc{1}{\beta}\pdr{\Phi_i(m_i(q),q)}{q}=\pdr{H_i(m_i(q),q)}{q}=m_i(q).
\]
Therefore, the matching condition
\be\label{26oct2512}
m_0(q)=m_1(q),
\ee
is equivalent to
\[
%m_0(q)=m_1(q)
%\quad\Longleftrightarrow\quad
(f_1-f_0)'(q)=0.
\]
Thus the matching values of $q$ that satisfy~\eqref{26oct2512} are precisely the critical points of the
free-energy difference $f_1-f_0$.

According to \cite{EPR-LMP}, in contact-geometric models of non-equilibrium thermodynamics
each such critical point corresponds to a Reeb chord
connecting the equilibrium Legendrian submanifolds of the two systems.
This phenomenon is of considerable interest, since Reeb chords are central objects of study in contact topology.
Let $q_*$ be such a critical point and write
\[
m_0(q_*)=m_1(q_*)=:m_*.
\]
In the thermodynamic limit, the instantaneous jump from system $0$ to system
$1$ sends an equilibrium macrostate to an equilibrium macrostate corresponding to
the same value $m_*$. Hence, the
transition is ultrafast. The same is true for the reverse jump.

By Remark~\ref{rem-free}, in this situation we have
\[
f_1(q_*)-f_0(q_*)
=
H_1(m_*,q_*)-H_0(m_*,q_*)
=
h_1(m_*)-h_0(m_*).
\]
Thus, in the thermodynamic limit, the average work per particle performed in going
along the Reeb chord is $f_1(q_*)-f_0(q_*)$, whereas the average work per particle
performed in going back is its negative. The two order-one contributions
cancel exactly in the complete cycle. In this sense, the cycle
along the chord and back becomes reversible
in the thermodynamic limit.

For a finite $N$, however, the two Gibbs measures need not coincide,
the average work over the cycle is strictly positive whenever they differ,
by Proposition~\ref{prop:work-positive}. At the matching value $q_*$,
expression~\eqref{26oct2718} gives
\[
w_N
=
\frac{(A_1-A_0)^2}{2\beta N A_0A_1}
+o(N^{-1}).
\]
Thus the Reeb-chord condition is characterized by the cancellation of the
order-one average work per particle performed along the chord and back, leaving only an
$O(N^{-1})$ finite-size contribution
(and possibly a smaller one when $A_0=A_1$).
This cancellation is the work-theoretic signature of an ultrafast process.

\noindent
{\bf Organization of the paper:}
In Section~\ref{sec-concentration-gibbs-measure}, we prove the key technical lemma
(Lemma~\ref{lem:26oct2704}) describing the concentration of the Gibbs measures
for the systems that we consider.
In Section~\ref{sec-asymptotics-of-moments}, we prove the asymptotics of the moments
that will be needed in the proof of Theorem~\ref{thm:main}.
In Section~\ref{sec-proof-main-theorem}, we prove Theorem~\ref{thm:main}.
We first establish the limiting distributions of the work and then use
bounds on the moments of the rescaled work to deduce the asymptotics
of its expectation and standard deviation.
In Section~\ref{sec:gaussian-example}, we consider
Gaussian oscillators -- an example of the class of systems
discussed in the previous sections. This section is short and technically
simple, and can be read independently of the rest of the paper.
Finally, in Appendix~\ref{app:large-dev}, we recall
two standard large deviation results used in the proof of
Theorem~\ref{thm:main}, and in Appendix~\ref{app:log-mgf-rate},
we prove some auxiliary facts about the log-moment generating function
and the rate function stated in Proposition~\ref{prop:log-mgf-rate} above.

 \noindent
 {\bf Acknowledgment:} L.P. thanks Kostya Khanin for an illuminating discussion at an early stage of this project. M.E.
was partially supported by the Israel Science
Foundation grant \# 2033/23. L.R. was supported by grants DMS-2205497 and DMS-2510166.

 \noindent
{\bf AI statement:}
We formulated and proved the result on the positivity of the
total average work discussed in Section~\ref{sec:cycle}.
Our original proof differs slightly from the one presented in the text:
the connection to the entropy was suggested by ChatGPT 6 Astra.
We discovered the central phenomenon of this paper and developed its proof in the context of
Gaussian oscillators, as described in Section~\ref{sec:gaussian-example}.
We also proposed the extension presented in Theorems~\ref{thm:main}, with certain technical assumptions subsequently supplied by ChatGPT~6~Astra. The proofs of these theorems, based on large deviation theory, were provided by ChatGPT~6~Astra. We then reworked and rewrote these proofs.

\section{Concentration of the Gibbs measures}\label{sec-concentration-gibbs-measure}

 Recall that by \eqref{26oct2602}, the Hamiltonians $NH_i(m)$ may have exponential growth at infinity.
The following concentration estimate will help us deal with this potential growth.

 \begin{lemma}
\label{lem:26oct2704}
Assume that the functions $H_i(m)$ satisfy the estimates~\eqref{26oct2320} and ~\eqref{26oct2602},
and $m_i$ is the unique minimizer of the function~$\Phi_i(m)$ defined in~\eqref{26oct2302}, and that $m_i$ is non-degenerate
in the sense of~\eqref{26oct2912}.
Then, for all~$r>0$ and
$\kappa\geq0$, there exist~$c_1 (r,\kappa)>0$ and~$c_2 (r,\kappa)>0$  and $N_0(r,\kappa)\in\N$ such that for $i=0,1$
and all $N\geq N_0(r,\kappa)$, we have
\begin{equation}\label{26oct2723}
I_{N,r,i}:=\int_{|m-m_i|>r}
e^{\kappa|m|}\,d\nu_i^{(N)}(m)=
 \frac1{Z_{i,N}}\int_{|m-m_i|>r}
e^{\kappa|m|} e^{-\beta N H_i(m)}\,P_N(dm)\le c_1e^{-c_2N}.
\end{equation}
 \end{lemma}

\begin{proof}
 Let us fix some $R>r$ and consider the intermediate integral
\be\label{26sep1410}
\bal
I_{med,N,i}&:=\int_{r\le|m-m_i|\le R}
e^{\kappa|m|}\,d\nu_i^{(N)}(m)
%=\farc{1}{Z_{i,N}}\int_{r\le |m-m_i|\le R}
%e^{\kappa|m|} e^{-\beta N H_i(m)}\,P_N(dm)\\
&\le
e^{\kappa (R+|m_i|)} \int_{r\le|m-m_i|\le R}
\,d\nu_i^{(N)}(m).
%=\farc{1}{Z_{i,N}}\int_{r\le |m-m_i|\le R}
%e^{-\beta N H_i(m)}\,P_N(dm).
\enbal
\end{equation}
As the function $H_i(m)$ is bounded from below by \eqref{26oct2320},
we may use the exponential-tilting corollary \eqref{26oct2930} of Varadhan's lemma,
Theorem~\ref{thm-lem-varadhan} below.
It shows that the probability measures
\be
d\nu_i^{(N)}=\farc{1}{Z_{i,N}}e^{-\beta N H_i(m)}\,P_N(dm),
\ee
defined by~\eqref{26oct2914},
satisfy a large deviation principle with the rate function
\[
J_i(m)=I(m)+\beta H_i(m)-\inf_{m\in\Rm}(I(m)+\beta H_i(m))
=\Phi_i(m)-\Phi_i(m_i).
\]
We deduce from \eqref{26oct2930} that
\be\label{26sep2618}
\limsup_{N\to+\infty}\farc{1}{N}\log(I_{med,N,i})\le -\inf_{r\le|m-m_i|\le R}J_i(m).
\ee
%Here, $I(m)$ is the rate function for the measures $P_N(dm)$.
Since $m_i$ is the unique minimizer of $\Phi_i(m)$, there exists $C_{r,R}$ such that
\be\label{26sep2620}
J_i(m)\ge C_{r,R}>0,~~\hbox{ for all $r\le|m-m_i|\le R$.}
\ee
Now,~\eqref{26sep2618} and~\eqref{26sep2620}  imply that
\be\label{26sep1420}
\limsup_{N\to+\infty}\farc{1}{N}\log(I_{med,N,i})\le -C_{r,R}.
\ee
%gives exponential decay on every compact subset of $\mathbb R\setminus V$.

Next, we control the  outer integral
\be
\bal
I_{out,N,i}&:=\farc{1}{Z_{i,N}}\int_{|m-m_i|\ge R }
e^{\kappa|m|} e^{-\beta N H_i(m)}\,P_N(dm)
\le \farc{e^{\beta ND_i}}{Z_{i,N}}\int_{|m-m_i|\ge R }
e^{\kappa|m|}  \,P_N(dm),
%\\
%&\le
%e^{\kappa (R+|m_i|)}\farc{1}{Z_{i,N}}\int_{r\le |m-m_i|\le R}
%e^{-\beta N H_i(m)}\,P_N(dm).
\enbal
\end{equation}
with the constant $D_i$ as in the assumption~\eqref{26oct2320} on $H_i(m)$.
%Here, we have set
%\be
%h_i=\inf_{m\in\Rm}H_i(m),
%\ee
%which is finite,
%\color{blue} since, by assumption (A), $H_i$ is bounded from below.
Note that for any $\eta>0$ fixed,  after increasing $R$ so that $R\geq4\max_i|m_i|$,  for all
$N\geq4\kappa/\eta$ we have
\be\label{26sep1412}
e^{\kappa|m|}
\leq e^{-\eta NR/2}e^{\eta N|m|},~~\hbox{ for all $m$ s.t. $|m-m_i|\ge R$.}
\ee
This gives
\be\label{26sep1414}
\bal
I_{out,N,i}&\le \farc{e^{-\eta NR/2}e^{\beta ND_i}}{Z_{i,N}}\int_{|m-m_i|\ge R }
e^{\eta N|m|} \,P_N(dm)\le
\farc{e^{-\eta NR/2}e^{\beta ND_i}}{Z_{i,N}}\int_{\Rm }
e^{\eta N|m|} \,P_N(dm).
\enbal
\ee
As we have assumed that the set $D$ defined by~\eqref{26sep1408} is an open interval,
we can choose~$\eta>0$ sufficiently small, so that
both~$\Lambda(\eta)$ and $\Lambda(-\eta)$ are finite, and set
\be\label{26sep1504}
M_\eta=\int_{\mathbb R}e^{\eta|s|}\rho(s)\,ds<+\infty.
\ee
%Here finiteness follows from
%$e^{\eta|s|}\leq e^{\eta s}+e^{-\eta s}$.
Recall that $m$ is related to the variables $s_j$ by~\eqref{26oct2812}, which implies
\be\label{26sep1502}
N|m|\leq\sum_{j=1}^N|s_j|.
\ee
It follows from the definition of the measure $P_N(dm)$
as the push-forward measure on the empirical mean $m$
and~\eqref{26sep1504}
and~\eqref{26sep1502}, that
\be\label{26sep1416}
\bal
\int_{\Rm }
e^{\eta N|m|}  \,P_N(dm)&=
\int_{\mathbb{R}^N} \exp \Big( \eta N \Big| \frac{1}{N} \sum_{j=1}^N s_j \Big|\Big)
\prod_{j=1}^N \rho(s_j) \, ds_j\\
&\le \int_{\mathbb{R}^N} \Big( \prod_{j=1}^N e^{\eta |s_j|} \Big)
\Big(\prod_{j=1}^N \rho(s_j) \Big)\, ds_1 \dots ds_N
= M_\eta^N.
\enbal
\ee

Finally, to bound $Z_{i,N}$, we note, once again that,
as the function $H_i(m)$ is bounded from below by \eqref{26oct2320},
Varadhan's lemma, Theorem~\ref{thm-lem-varadhan},
can be applied to the function $(-\beta H_i(m))$. This gives
\be
\label{26oct2621}
\bal
 \lim_{N\to +\infty}&\frac1N\log Z_{i,N}=\lim_{N\to +\infty}\frac1N\log
 \int_{\mathbb R}e^{-\beta NH_i(m)}\,P_N(dm)
 \\
 &=
 \sup_{m\in\mathbb R}
 \{-\beta H_i(m)-I(m)\}=\sup_{m\in\Rm}(-\Phi_i(m))=-\inf_{m\in\Rm}\Phi_i(m)=-\Phi_i(m_i).
\enbal
\ee
Thus, for every $\varepsilon>0$ and all
sufficiently large $N$, we have
\be\label{26sep1418}
 Z_{i,N}\geq e^{-N(\Phi_i(m_i)+\varepsilon)}.
\ee
Putting together~\eqref{26sep1414},~\eqref{26sep1416}, and~\eqref{26sep1418}, we get that
there exist $c_1,c_2>0$ such that
\be
\bal
I_{out,N,i}&\le \farc{e^{-\eta NR/2}e^{ \beta ND_i}}{Z_{i,N}} \int_{\Rm }
e^{\eta N|m|} \,P_N(dm)
 \\
 &\leq
 \exp\Big\{-N\Big[
 \frac{\eta R}{2}-\log M_\eta-\Phi_i(m_i)-\varepsilon-\beta D_i
 \Big]\Big\}\le c_1e^{-c_2N},
\enbal
\ee
if we choose $R$ sufficiently large. Together with~\eqref{26sep1420}, this proves
\eqref{26oct2723}
and thus finishes the proof of the lemma.
\end{proof}

 \section{Asymptotics of the moments}
\label{sec-asymptotics-of-moments}

In this section, we prove the following asymptotics of the moments
that will be needed in the proof of Theorem~\ref{thm:main}. We set
\be\label{26oct2623}
\alpha_i:=\frac{\beta H_i''(m_i)}{2},
\ee
and
\be\label{26oct2633}
\sigma_i^2:=\Lambda''(\theta_i).
\ee
Recalling the definition~\eqref{26oct2912} of $A_i$, and using~\eqref{26oct2629}
and~\eqref{26oct2316}, we see that
\begin{equation}\label{26oct2627}
A_i:=\Phi_i''(m_i)=I''(m_i)+\beta H_i''(m_i)
=\frac{1}{\sigma_i^2}+2\alpha_i.
\end{equation}

\begin{lemma}[Asymptotics of the moments]\label{lem-26oct2602}
For any $k\in\Zm_{\ge 0}$ we have
\be\label{26sep2622}
\bal
N^{k/2}\int_{\mathbb R}(m-m_i)^{k}\,d\nu_i^{(N)}(m)\to
\farc{\Em[\xi^{k}e^{-{\alpha_i}\xi^2}]}{\Em[e^{-{\alpha_i}\xi^2}]},~~\hbox{as $N\to+\infty$}.
\enbal
\ee
Here, $\xi$ is a
Gaussian random variable
with mean $0$ and variance $\sigma_i^2$.
\end{lemma}
The rest of this section contains the proof of Lemma~\ref{lem-26oct2602}.
The plan of the proof is as follows. We will prove \eqref{26sep2622} by reducing it to the central
limit theorem for a certain collection of random variables. To this end, we will rewrite
in Section~\ref{sec:tilted} the left
side of \eqref{26sep2622} in a different form in \eqref{26sep2310}.
Then, in Section~\ref{sec:second-moment}, we will use \eqref{26sep2310} to prove \eqref{26oct2634},
an equivalent form of \eqref{26sep2622}. This will be done modulo a tail
estimate \eqref{26oct3010} whose proof will be given
in Section~\ref{subsec-proof-of-tail-estimate}.

We will now make a few preparations for the proof of Lemma~\ref{lem-26oct2602}.
We drop the subscript~$i$ throughout the proof
(in particular, in $\alpha_i$, $\sigma_i$)
and denote
\be
\hbox{$m_* :=m_i$, $A_* := A_i$ and $\theta_* :=\theta_i$.}
\ee
For an odd $k=2p-1$, the right side of~\eqref{26sep2622} equals zero.
For even $k=2p$, we may iteratively compute, using \eqref{26oct2627} and integration by parts:
%(we suppress the index $i$ in $\alpha_i$, $\sigma_i$, and $A_i$)
\be\label{26oct2625}
\bal
I_p:&=\Em\big[\xi^{2p}e^{-\alpha \xi^2}\big]=
\int_\Rm z^{2p}\farc{e^{-z^2/(2\sigma^2)}}{(2\pi\sigma^2)^{1/2}}e^{-\alpha z^2}dz
=\farc{1}{(2\pi\sigma^2)^{1/2}}
\int_\Rm z^{2p}e^{-A_* z^2/2}dz\\
&=
\farc{1}{A_*^{(2p+1)/2}}\farc{1}{(2\pi\sigma^2)^{1/2}}
\int_\Rm z^{2p}e^{-z^2/2}dz=\farc{(2p-1)}{A_*^{(2p+1)/2}}\farc{1}{(2\pi\sigma^2)^{1/2}}
\int_\Rm z^{2p-2}e^{-z^2/2}dz\\
&=\farc{2p-1}{A_*}\farc{1}{A_*^{(2(p-1)+1)/2}}\farc{1}{(2\pi\sigma^2)^{1/2}}
\int_\Rm z^{2(p-1)}e^{-z^2/2}dz
=\farc{2p-1}{A_*}I_{p-1}.
\enbal
\ee
We conclude that
\be
\farc{I_p}{I_0}=\farc{(2p-1)\cdot(2p-3)\cdot\dots\cdot 3\cdot 1}{A_*^p}.
\ee
In particular, for $k=2$ we have
\be\label{26sep2630}
\bal
N\int_{\mathbb R}(m-m_*)^{2}\,d\nu^{(N)}(m)\to \farc{1}{A_*},~~\hbox{as $N\to+\infty$},
\enbal
\ee
and for $k=4$ we have
\be\label{26oct2737}
N^2\int_{\mathbb R}(m-m_*)^{4}\,d\nu^{(N)}(m)\to \farc{3}{A_*^2},~~\hbox{as $N\to+\infty$}.
\ee
%In the proof of Lemma~\ref{lem-26oct2602} we drop the subscript $i$ throughout the proof and denote
%\be
%\hbox{$m_*=m_i$, $A_i=A_*$ and $\theta_*=\theta_i$.}
%\ee

 \subsection{The exponentially tilted density}
\label{sec:tilted}

 We start proving~\eqref{26sep2622} by reducing it to the central limit theorem for a certain
collection of random variables.  In this preliminary section, we will rewrite the left side of~\eqref{26sep2622}
in the form
\be\label{26sep2310}
N^{k/2}\int_{\mathbb R}(m-m_*)^k\,d\nu^{(N)}(m)
=\farc{\mathbb E_{\theta_*}\big[{U_N^k}e^{-R_N(U_N)}\big]}{\mathbb E_{\theta_*}\big[ e^{-R_N(U_N)}\big]
} .
\ee
More generally, we will show that for any function $F(m)$ that is integrable with
respect to the measure $\nu^{(N)}(m)$, we have
\begin{equation} \label{26sep2312}
 \int_{\mathbb R}F(m)\,d\nu^{(N)}(m)
 =\farc{1}{\mathbb E_{\theta_*}\big[ e^{-R_N(U_N)} \big]}\mathbb E_{\theta_*}\Big[
 F\Big(m_*+U_N/\sqrt N\Big)e^{-R_N(U_N)}
 \Big].
\ee
In the rest of Section~\ref{sec:tilted}, we explain the notions that appear in the right side
of~\eqref{26sep2310} and~\eqref{26sep2312} and derive these identities.

We define the exponentially tilted
density
\[
\rho_{\theta_*}(x)=e^{\theta_* x-\Lambda(\theta_*)}\rho(x),~~x\in\Rm.
\]
Note that $\rho_{\theta_*}(x)$ is a probability density by the
definition~\eqref{26oct2932} of $\Lambda(\theta_*)$. Given a collection of
independent random variables $X_1,\dots,X_N$, each distributed according to $\rho_{\theta_*}(x)$,
we will denote by~$\mathbb E_{\theta_*}$ the expectation with respect to the joint
law
of $X_j$, $j=1,\dots,N$:
\be
\Em_{\theta_*}[f(X_1,\dots,X_N)]=\int_{\Rm^N}f(x)\rho_{\theta_*}(x_1)\dots\rho_{\theta_*}(x_N)dx,
\ee
for any function $f(x)$ integrable with respect to the product measure in the above integral.

Note that the mean of each $X_j$ is
\[
\mathbb E_{\theta_*} X_j=\int_\Rm x \rho_{\theta_*}(x)dx=
\int_\Rm s e^{\theta_* s-\Lambda(\theta_*)}\rho(s)ds
=e^{-\Lambda(\theta_*)}\farc{d}{d\theta}\Big(\int_\Rm e^{\theta s}\rho(s)ds\Big)
\Big|_{\theta=\theta_*}=\Lambda'(\theta_*).
\]
Since, by \eqref{26oct2502}, we have
\be\label{26oct1802}
\Lambda'(\theta_*) = m_*,
\ee
we get
\be\label{26sep2308}
\mathbb E_{\theta_*} X_j=
m_*.
\ee
Recalling the
definition~\eqref{26oct2932} of $\Lambda(\theta)$
% and using again the equality $\Lambda'(\theta_*) = m_*$,
%the relation~\eqref{eq:scalar-equilibrium-equations} between~$\Lambda'(\theta)$ and $m_*$,
we can compute
the corresponding variance as
\be
\bal
\operatorname{Var}_{\theta_*}(X_j)&=\int_\Rm (x-m_*)^2\rho_{\theta_*}(x)dx=
\int_\Rm (s-m_*)^2 e^{\theta_* s-\Lambda(\theta_*)}\rho(s)ds\\
&
= e^{-\Lambda(\theta_*)}\int_\Rm(s^2-2m_*s+m_*^2)e^{\theta_* s}\rho(s)ds \\
&=
e^{-\Lambda(\theta_*)}
\Big[\farc{d^2}{d\theta^2}-2m_*\farc{d}{d\theta}+m_*^2\Big]
\Big(\int_\Rm e^{\theta s}\rho(s)ds\Big)\Big|_{\theta=\theta_*}
\\
&=e^{-\Lambda(\theta)}
\Big[\farc{d^2}{d\theta^2}-2m_*\farc{d}{d\theta}+m_*^2\Big]e^{\Lambda(\theta)}\Big|_{\theta=\theta_*}.
\enbal
\ee
Together with~\eqref{26oct1802}, this yields
\be\label{26sep1538}
\operatorname{Var}_{\theta_*} (X_j) =\Lambda''(\theta_*)+|\Lambda'(\theta_*)|^2-2m_*\Lambda'(\theta_*)+m_*^2
=\Lambda''(\theta_*)=\sigma^2,
\ee
due to~\eqref{26oct2633}.

%Since $P_N$ is the push-forward of the product density
%$\prod_{j=1}^N\rho(x_j)$ under the map
%\[
% (x_1,\ldots,x_N)\longmapsto
% \overline x:=\frac1N\sum_{j=1}^Nx_j,
%\]
%we have,   for every function $F$
%integrable with respect to $\nu_i^{(N)}$,
%\be\label{26sep1423}
%\bal
%&\int_{\mathbb R}F(m)e^{-\beta NH(m)}P_N(dm)
%=\int_{\mathbb R^N}
%F(\overline x)e^{-\beta NH(\overline x)}
%\prod_{j=1}^N\rho(x_j)\,dx_1\cdots dx_N
%\\
%&
% =
% e^{N\Lambda(\theta)}
% \int_{\mathbb R^N}
% F(\overline x)e^{-NG_\theta(\overline x)}
% \prod_{j=1}^N\rho_\theta(x_j)\,dx_1\cdots dx_N
% =
% e^{N\Lambda(\theta)}
% \mathbb E_\theta
% \left[
% F(\overline X_N)e^{-NG_\theta(\overline X_N)}
% \right],
%\enbal
%\ee
%with the function
%\be\label{26sep1522}
%G_\theta(m):=\beta H(m)+\theta m.
%\ee

%Using the empirical mean
%\be
%\overline X_N:=\farc{1}{N}\sum_{j=1}^NX_j,
%\ee
%we may represent the joint density of the variables $S_k$ as
%\be\label{26sep1512}
% \prod_{j=1}^N\rho(x_j)
% =
% e^{-N\theta\overline x+N\Lambda(\theta)}
% \prod_{j=1}^N\rho_\theta(x_j).
%\ee
%\color{red}[ME: $\overline x$ is used before it is defined].\color{black}

Keeping in mind~\eqref{26sep2308}, it is convenient to introduce the centered and
rescaled empirical mean, in the spirit of the central limit theorem,
\be\label{26oct710}
 U_N=\frac1{\sqrt N}\sum_{j=1}^N(X_j-m_*),
\ee
as well as
\be\label{26oct1820}
\bar X_N=\farc{1}{N}\sum_{k=1}^NX_k.
\ee
%so that
%\[
% \overline X_N=m_*+\frac{U_N}{\sqrt N}.
%\]
We also define the function
\be\label{26sep1522}
G(m):=\beta H(m)+\theta_*m,
\ee
and set
\be\label{26sep1520}
 R_N(u)=N\left[
 G\left(m_*+\frac{u}{\sqrt N}\right)-G(m_*)
 \right],
\ee
so that
\be\label{26oct702}
NG (\overline X_N)=NG (m_*)+R_N(U_N).
\ee

Since $P_N$ is the push-forward of the product density
$\prod_{j=1}^N\rho(x_j)$ under the map
\[
 (x_1,\ldots,x_N)\longmapsto
 \overline x:=\frac1N\sum_{j=1}^Nx_j,
\]
we have,   for every function $F$
integrable with respect to $\nu^{(N)}$,
\be\label{26sep1423}
\bal
&\int_{\mathbb R}F(m)e^{-\beta NH(m)}P_N(dm)
=\int_{\mathbb R^N}
F(\overline x)e^{-\beta NH(\overline x)}
\prod_{j=1}^N\rho(x_j)\,dx_1\cdots dx_N
\\
&
 =
 e^{N\Lambda(\theta_*)}
 \int_{\mathbb R^N}
 F(\overline x)e^{-NG (\overline x)}
 \prod_{j=1}^N\rho_{\theta_*}(x_j)\,dx_1\cdots dx_N
 =
 e^{N\Lambda(\theta_*)}
 \mathbb E_{\theta_*}
 \left[
 F(\overline X_N)e^{-NG(\overline X_N)}
 \right].
\enbal
\ee
Inserting expressions~\eqref{26sep1522}-\eqref{26oct702} into the right-hand side of~\eqref{26sep1423},
we obtain
\be\label{26sep1424}
 \int_{\mathbb R}
 F(m)e^{-\beta NH(m)}P_N(dm)
 =
 e^{N\Lambda(\theta_*)-NG(m_*)}
 \mathbb E_{\theta_*}
 \Big[
 F\Big(m_*+\frac{U_N}{\sqrt N}\Big)
 e^{-R_N(U_N)}
 \Big].
\ee
Taking, in particular, $F=1$ in~\eqref{26sep1424} gives
\be\label{26sep1425}
Z_{N}=e^{N\Lambda(\theta_*)-NG  (m_*)}
 \mathbb E_{\theta_*}\Big[ e^{-R_N(U_N)}\Big].
\ee
Dividing these two identities and recalling \eqref{26oct2914}
yields
\begin{equation} \label{26oct3006}
 \int_{\mathbb R}F(m)\,d\nu^{(N)}(m)
 =\farc{1}{\mathbb E_{\theta_*}\Big[ e^{-R_N(U_N)}\Big]}\mathbb E_{\theta_*}\Big[
 F\Big(m_*+\farc{U_N}{\sqrt N}\Big)e^{-R_N(U_N)}
 \Big].
\ee
%with
%\be\label{26oct704}
%D_N=\mathbb E_\theta\Big[ e^{-R_N(U_N)}\Big].
%\end{equation}
%Together,~\eqref{26oct3006}-\eqref{26oct704} represent~\eqref{26sep2312}.
This is~\eqref{26sep2312}.
Note that the left side of~\eqref{26oct3006} with $F(m)=(m-m_*)^{k}$
is exactly the integral in the left side of~\eqref{26sep2622} that we need to estimate
in Lemma~\ref{lem-26oct2602}:
\be\label{26sep1418-1}
\bal
\int_{\mathbb R}(m-m_*)^{k}\,d\nu^{(N)}(m)
& =\frac{1}{\mathbb E_{\theta_*}\Big[ e^{-R_N(U_N)}\Big]}
\mathbb E_{\theta_*}\Big[
\farc{U_N^{k}}{N^{k/2}}e^{-R_N(U_N)}\Big]
\\
&=\farc{1}{N^{k/2}\mathbb E_{\theta_*}\Big[ e^{-R_N(U_N)}\Big]}
\mathbb E_{\theta_*}\Big[{U_N^k}e^{-R_N(U_N)}
 \Big],
\enbal
\ee
which is~\eqref{26sep2310}.

This finishes the proof of
\eqref{26sep2310} and \eqref{26sep2312}.

 \subsection{The weighted moments}
\label{sec:second-moment}

With~\eqref{26sep2310} in hand, in order to prove~\eqref{26sep2622}, we need to show that
\begin{equation}\label{26oct2634}
\bal
\frac{
\mathbb E_{\theta_*}\big[U_N^ke^{-R_N(U_N)}\big]
 }{
 \mathbb E_{\theta_*}\big[ e^{-R_N(U_N)}\big]
 }\to
\farc{\Em[\xi^{k}e^{-\alpha \xi^2}]}{\Em[e^{-\alpha \xi^2}]},~~\hbox{as $N\to+\infty$}.
\enbal
\end{equation}
Here, $\xi$ is a $\cN(0,\sigma^2)$ normal variable.
We first note that, as $\theta_*\in\hbox{int}(D)$, all moments of $X_k$ are finite:
\be\label{26oct3008}
\Em_{\theta_*}(X_k^{2p})=\int_{\Rm}x^{2p}\rho_{\theta_*} (x)dx=
\int_{\Rm} s^{2p}e^{\theta_* s-\Lambda(\theta_*)}\rho(s) ds<+\infty,
\ee
for all $p\in\Nm$.
Thus, the central limit theorem, together with~\eqref{26sep1538}, implies that
\be\label{26sep1525}
 U_N\Longrightarrow \xi,
 \qquad
 \xi\sim\mathcal N(0,\sigma^2).
\ee
All moments of $U_N$ are bounded by \eqref{26oct3008},
independence and centering:
\[
\sup_{N\ge1}\mathbb E_{\theta_*}|U_N|^{2r}<\infty,\ \textrm{for all}\ r\in\N.
\]
Hence, \eqref{26sep1525} yields, for every bounded continuous $g$ with
$\mathbb E g(\xi)\ne0$,
\be
\frac{\Em_{\theta_*}\left[U_N^kg(U_N)\right]}{\Em_{\theta_*}[g(U_N)]}\to \farc{\Em(\xi^kg(\xi))}{\Em g(\xi)}.
\ee
The only difficulty in analyzing the left side of~\eqref{26oct2634}
is that the function $\exp(-R_N(u))$ defined by~\eqref{26sep1520} is not necessarily bounded in $u$,
and depends on $N$.
On the other hand, this exponential is not far from being
both bounded in $u$ and nearly independent of $N$, since it
converges uniformly on compact sets to
$e^{-\alpha u^2}$, as $N\to +\infty$.
Indeed, the function~$G (m)$ defined
by~\eqref{26sep1522} satisfies
\be\label{26sep1523}
G'(m_*):=\beta H'(m_*)+\theta_*=0,
\ee
by~\eqref{26oct2635}.
As $G''(m)=\beta H''(m)$,  Taylor’s
theorem gives
%there exist~$N_0$ and~$C_K>0$  so that for all $N\geq N_0$
\be\label{26sep1521}
\bal
R_N(u)&=N\Big[
 G\Big(m_*+\frac{u}{\sqrt N}\Big)-G (m_*)
 \Big]=\farc{\beta H''(m_*)}{2} u^2+r_N(u),
\enbal
\ee
with
\be
\bal
\sup_{u\in K}|r_N(u)|\to 0,~~\hbox{as $N\to+\infty$,}
%\le\farc{C_K}{\sqrt{N}} ~~\hbox{for all $u\in K$ and $N\ge N_0$.}
\enbal
\ee
for any compact set $K\subset\Rm$.

However, at this stage we only know that $r_N$ converges uniformly to zero on compact
sets, as $N\to +\infty$ and not on all of $\R$.
\color{black}
%while the $o(1)$ term in~\eqref{26sep1521} is uniformly in $N$
%bounded on compact sets, it is not
%necessarily bounded on all of $\Rm$ uniformly in $N$.
To circumvent this problem, we introduce a smooth cut-off
function~$\chi_M(u)$  with an integer $M\geq1$ such that
\[
 \chi_M(u)=1 \quad\text{for } |u|\leq M,
 \qquad
 \chi_M(u)=0 \quad\text{for } |u|\geq2M,
\]
and $0\le \chi_M(u)\le 1$ for all $u\in\Rm$.  Then,~\eqref{26sep1521}
and the central limit theorem~\eqref{26sep1525} imply that   we have
\be\label{26sep1524}
 \mathbb E_{\theta_*}\left[
 \chi_M(U_N)U_N^k e^{-R_N(U_N)}
 \right]
 \longrightarrow
 \mathbb E\left[
 \chi_M(\xi)\xi^k e^{-\alpha \xi^2}
 \right],~~\hbox{as $N\to+\infty$,}
\ee
with $\alpha$ given by~\eqref{26oct2623}.
%\be\label{26oct1202}
%\alpha=\frac{\beta H''(m_*)}{2}.
%\ee
Note that $\alpha$ is exactly the coefficient in front of $u^2$ in \eqref{26sep1521}.

To remove the cut-off, it remains to prove that
we have the following tail estimate
\begin{equation} \label{26oct3010}
 \lim_{M\to +\infty}\limsup_{N\to +\infty}
 \mathbb E_{\theta_*}\left[|U_N|^k e^{-R_N(U_N)};
 |U_N|>M
 \right]
 =0.
\end{equation}
Applying \eqref{26oct3010} with exponents $0$ and $k$, we
%Assuming this estimate, given any $\eps>0$, we first
choose $M_\eps$ sufficiently large,
so that
\begin{equation} \label{26sep1526}
 \limsup_{N\to +\infty}
 \mathbb E_{\theta_*}\left[
 (1+|U_N|^k) e^{-R_N(U_N)};
 |U_N|>M
 \right]<\eps,~~\hbox{ for all $M>M_\eps$},
\end{equation}
and then, for all $M>M_\eps$ pick $N_\eps(M)$ sufficiently large, so that
\begin{equation} \label{26sep1527}
 \mathbb E_{\theta_*}\left[
 (1+|U_N|^k) e^{-R_N(U_N)};
 |U_N|>M
 \right]<2\eps,~~\hbox{ for all $M>M_\eps$ and $N>N_\eps(M)$.}
\end{equation}
Next, for any $M\ge 1$ and $N\ge 1$, we decompose
\be\label{26sep1528}
\bal
%D_N&=
\mathbb E_{\theta_*}\big[ e^{-R_N(U_N)}\big]
&=\mathbb E_{\theta_*}\left[\chi_M(U_N)e^{-R_N(U_N)}
 \right]+\mathbb E_{\theta_*}\left[(1-\chi_M(U_N))e^{-R_N(U_N)}
 \right]
 \\
 &=I_{M,N}+II_{M,N}.
\enbal
\ee
For the first term, we have, due to~\eqref{26sep1524},
\be
I_{M,N}\to \mathbb E\big[
 \chi_M(\xi)e^{-\alpha \xi^2}
 \big],~~\hbox{as $N\to+\infty$, for all $M\ge 1$,}
\ee
so that there is $N_\eps'(M)$ such that
\be\label{26sep1530}
\Big|I_{M,N}- \mathbb E\big[
 \chi_M(\xi)e^{-\alpha \xi^2}
 \big]\Big|<\eps,~~\hbox{ for all $N>N_\eps'(M)$.}
\ee
In addition, we can find $M_\eps'$ so that
\be\label{26sep1531}
\Big|\mathbb E\big[
 \chi_M(\xi)e^{-\alpha \xi^2}
 \big]- \mathbb E\big[e^{-\alpha \xi^2}
 \big]\Big|<\eps\,~~\hbox{for all $M>M_\eps'$.}
\ee
The second term in the right side of~\eqref{26sep1528} can be bounded using~\eqref{26sep1527}:
\be\label{26sep1532}
II_{M,N}\le \mathbb E_{\theta_*}\big[e^{-R_N(U_N)};
 |U_N|>M
 \big]<2\eps,~~\hbox{ for all $M>M_\eps$ and $N>N_\eps(M)$.}
\ee
Given $\eps>0$, we choose $M>\max(M_\eps,M_\eps')$  and
$N>\max(N_\eps(M),N_\eps'(M))$, and write
\be
\bal
\big|\mathbb E_{\theta_*}\big[ e^{-R_N(U_N)}\big]- \mathbb E\big[e^{-\alpha \xi^2}
 \big]\big|&\le \big|I_{M,N}-\Em\big[\chi_M(\xi)e^{-\alpha \xi^2}\big]\big|+\Big|\mathbb E\big[
 \chi_M(\xi)e^{-\alpha \xi^2}
 \big]- \mathbb E\big[e^{-\alpha \xi^2}
 \big]\Big|\\
 &+II_{M,N}\le 20\eps,
%=\mathbb E_\theta\big[ e^{-R_N(U_N)}\big]
%=\mathbb E_\theta\left[\chi_M(U_N)e^{-R_N(U_N)}
% \right]+\mathbb E_\theta\left[(1-\chi_M(U_N))e^{-R_N(U_N)}
% \right]
% \longrightarrow\mathbb E e^{-\alpha \xi^2}>0
\enbal
\ee
due to~\eqref{26sep1530},~\eqref{26sep1531} and~\eqref{26sep1532}.

A nearly identical argument shows that
\[
 \mathbb E_{\theta_*}\left[U_N^ke^{-R_N(U_N)}\right]
 \longrightarrow
 \mathbb E\big[\xi^ke^{-\alpha \xi^2}\big].
\]
We see that
\begin{equation}\label{26sep1535}
\frac{
\mathbb E_{\theta_*}\big[U_N^ke^{-R_N(U_N)}\big]
 }{
 \mathbb E_{\theta_*}\big[ e^{-R_N(U_N)}\big]
 }
 \to\farc{\Em\big[\xi^ke^{-\alpha \xi^2}\big]}{\mathbb E\big[e^{-\alpha \xi^2}\big]},
\end{equation}
which is~\eqref{26oct2634}.
%Recall that $\xi$ is a mean-zero real-valued Gaussian variable with variance $\sigma$ and that,
%because of~\eqref{26oct1204} and~\eqref{26oct1202}, we have
%\be\label{26oct2037}
%\frac1{2\sigma^2}+\alpha=\frac A2.
%\ee
%We may now explicitly compute
%\be\label{26oct2034}
%\bal
%\Em\big[e^{-\alpha \xi^2}\big]=\int_\Rm \farc{e^{-z^2/(2\sigma^2)}}{(2\pi\sigma^2)^{1/2}}e^{-\alpha z^2}dz=\farc{1}{(2\pi\sigma^2)^{1/2}}
%\int_\Rm e^{-Az^2/2}dz=\farc{1}{\sqrt{A}}\farc{1}{(2\pi\sigma^2)^{1/2}}
%\int_\Rm e^{-z^2/2}dz,
%\enbal
%\ee
%and
%\be\label{26oct2035}
%\bal
%\Em\big[\xi^2e^{-\alpha \xi^2}\big]&=\int_\Rm z^2\farc{e^{-z^2/(2\sigma^2)}}{(2\pi\sigma^2)^{1/2}}e^{-\alpha z^2}dz=\farc{1}{(2\pi\sigma^2)^{1/2}}
%\int_\Rm z^2e^{-Az^2/2}dz\\
%&=
%\farc{1}{A^{3/2}}\farc{1}{(2\pi\sigma^2)^{1/2}}
%\int_\Rm z^2e^{-z^2/2}dz=\farc{1}{A^{3/2}}\farc{1}{(2\pi\sigma^2)^{1/2}}
%\int_\Rm e^{-z^2/2}dz.
%\enbal
%\ee
%%
%%Moreover, multiplying the density
%%of $\xi$ by $e^{-\alpha z^2}$ and normalizing gives the Gaussian density
%%proportional to
%%\[
%% \exp\Big\{-\frac A2z^2\Big\}.
%%\]
%%It is the density of $\mathcal N(0,A^{-1})$.
%Now,~\eqref{26sep1535}
%gives~\eqref{26sep2622}.

\subsection{The proof of the tail estimate}
\label{subsec-proof-of-tail-estimate}

Thus, to finish the proof of Lemma~\ref{lem-26oct2602}
it remains only to prove the tail estimate
\eqref{26oct3010}. We first claim the following.

 \begin{lemma}\label{lem-26oct1802}
For any $\eps>0$ there exists $C_\eps>0$ so that
for all $N\in\N$,  we have
\begin{equation}\label{26oct1806}
\Em_{\theta_*}\Big[e^{-R_N(U_N)}\Big]\le C_\eps e^{\eps N}.
\ee
\end{lemma}

\begin{proof} Varadhan's lemma (Theorem~\ref{thm-lem-varadhan} in the appendix) implies that
\be\label{26oct1824}
\lim_{N\to+\infty}\frac{1}{N}\log\Big(\int_{\Rm}e^{-\beta NH(m)}P_N(dm)\Big)=\sup_{m\in\Rm}\big[-\beta H(m)-I(m)\big]=\sup_{m\in\Rm}[-\Phi(m)]=-\Phi(m_*).
\ee
We also have
\be\label{26oct1821}
\bal
\Em_{\theta_*}\Big[e^{-R_N(U_N)}\Big]&=\Em_{\theta_*}\Big[\exp\Big(-N\Big[ G\Big(m_*+\farc{U_N}{\sqrt{N}}\Big)-G(m_*)\Big]\Big)
\Big]\\
&=\Em_{\theta_*}\Big[\exp\Big(-N\Big[ \beta H\Big(m_*+\farc{U_N}{\sqrt{N}}\Big)-\beta H(m_*)+\farc{\theta_*U_N}{\sqrt{N}}\Big]\Big)\Big]\\
&=\Em_{\theta_*}\Big[\exp\Big(-N\Big[ \beta H(\bar X_N)-\beta H(m_*)+\theta_*(\bar X_N-m_*)\Big]\Big)\Big].
\enbal
\ee
Here, we have passed back to the variables $\bar X_N$ defined in~\eqref{26oct1820}. Now, as
in~\eqref{26sep1423} but read backwards, and using~\eqref{26oct1821}
and the notation
\[
{\bar x}_N := N^{-1} \sum_{j=1}^N x_j,
\]
we have
\be\label{26oct1822}
\bal
\Em_{\theta_*}\Big[e^{-R_N(U_N)}\Big]&=\int_{\Rm^N} \exp\Big(-N\Big[ \beta H({\bar x}_N)-\beta H(m_*)+\theta_*({\bar x}_N-m_*)\Big]\Big)\\
&\times e^{\theta_*x_1+\dots+\theta_* x_N-N\Lambda(\theta_*)}\rho(x_1)\dots\rho(x_N)dx_1\dots dx_N\\
&=e^{-N\Lambda(\theta_*)+N\beta H(m_*)+N\theta_*m_*}\int_\Rm e^{-\beta NH(m)}P_N(dm)\\
&=e^{-N\Lambda(\theta_*)+N\Phi(m_*)-NI(m_*)+N\theta_*m_*}
\int_\Rm e^{-\beta NH(m)}P_N(dm).
\enbal
\ee
Furthermore, as the function $\Lambda(\theta)$ is strictly convex in $\theta$,
the function~$\theta m-\Lambda(\theta)$ attains at most one maximum,
and, because of~\eqref{26oct2314}, this maximum is attained at~$\theta=\theta_*$. That is, we have
\be
I(m_*)=\sup_{\theta\in\Rm}[\theta m_*-\Lambda(\theta)]=m_*\theta_*-\Lambda(\theta_*).
\ee
Using this identity in~\eqref{26oct1822}, we get
\be\label{26oct1823}
\bal
\Em_{\theta_*}\Big[e^{-R_N(U_N)}\Big]&=e^{ N\Phi(m_*) }
\int_\Rm e^{-\beta NH(m)}P_N(dm).
\enbal
\ee
Recalling~\eqref{26oct1824}, we arrive at
\be
\bal
\lim_{N\to+\infty}\farc{1}{N}\log\Em_{\theta_*}\Big[e^{-R_N(U_N)}\Big]=\Phi(m_*)+\lim_{N\to+\infty}\farc{1}{N}\log\Big[\int_\Rm e^{-\beta NH(m)}P_N(dm)\Big]=0.
\enbal
\ee
Now, \eqref{26oct1806} follows.
\end{proof}

We turn to the proof of the tail estimate
\eqref{26oct3010}.
Let us choose $\delta>0$, to be fixed later, and consider separately the events $\{|U_N|\ge \delta\sqrt{N}\}$
and $\{M\le |U_N|\le\delta\sqrt{N}\}$.
First, we look at
the event~$\{|U_N|\ge \delta\sqrt{N}\}$.
%On the event $|U_N|>\delta\sqrt N$, the empirical mean is a fixed distance
%from $m_*$.
Applying the identity~\eqref{26oct3006} to the function
\[
 F_N(m)=N^{k/2}|m-m_*|^k
 \mathbf 1_{\{|m-m_*|>\delta\}},
\]
we get
\be\label{26sep1545}
\bal
\mathbb E_{\theta_*}\left[
 |U_N|^k e^{-R_N(U_N)};
 |U_N|>\delta\sqrt N
 \right]
  &=
\Em_{\theta_*}\Big[e^{-R_N(U_N)}\Big]N^{k/2}
 \int_{|m-m_*|>\delta}
 |m-m_*|^k\,d\nu^{(N)}(m)\\
 &\le c_1e^{-c_2N}.
\enbal
\ee
We used the bounds \eqref{26oct2723} and~\eqref{26oct1806}
in the last step above. Indeed, since
\[
|m-m_*|^k\le C_k e^{|m|},
\]
the estimate~\eqref{26oct2723} bounds the integral by $Ce^{-\gamma N}$ for some
$\gamma>0$. Next, we apply \eqref{26oct1806} with $\eps=\gamma/2$.
The remaining factor $N^{k/2}$ can then be absorbed by decreasing
the decay exponent, for instance to $\gamma/4$.

It remains to consider the event $\{M\le |U_N|\le\delta\sqrt{N}\}$.
Note that
\be
G'(m_*)=\beta H'(m_*)+\theta_*=0,
\ee
by~\eqref{26oct2635}, while
\be
G''(m_*)=\beta H''(m_*)=A_*-\farc{1}{\sigma^2},
\ee
by~\eqref{26oct2623} and~\eqref{26oct2627}.
%and $m_*$ is a strict local minimum of the function $G(m)$,
Fix $0<\eps<A_*$. We can choose $h_\eps>0$ so that
\be\label{26nov902}
G(m_*+h)-G(m_*)\ge \Big(A_*-\farc{1}{\sigma^2}-\eps\Big)\farc{h^2}{2},~~
\hbox{ for $|h|\le h_\eps$.}
\ee
Let us set
\be\label{26oct1814}
a_\eps=\max\Big\{0,\farc{1}{2\sigma^2}+\farc{\eps}{2}-\farc{A_*}{2}\Big\}=\farc{1}{2}
\max\Big\{0,\farc{1}{\sigma^2}+\eps-A_*\Big\}.
\ee
Our choice of $\eps$ ensures that
\be\label{26nov910}
0\le a_\eps<\farc{1}{2\sigma^2}.
\ee
It follows from~\eqref{26nov902} and~\eqref{26oct1814} that, as long as
\be\label{26nov1906}
{0<}\delta\le  h_\eps,
\ee
we have
\be\label{26sep1537}
-R_N(u)=-N\Big[
 G\Big(m_*+\frac{u}{\sqrt N}\Big)-G(m_*)
 \Big]\le a_\eps u^2,~~\hbox{ for $|u|\le\delta{\sqrt{N}}$.}
\ee
We now use~\eqref{26sep1537} to obtain
\be\label{26sep1539}
\bal
\mathbb E_{\theta_*}\Big[
 |U_N|^k e^{-R_N(U_N)};
 &~M\le |U_N|\le \delta\sqrt N
 \Big]
  \le \mathbb E_{\theta_*}\left[
 |U_N|^k e^{a_\eps U_N^2};
 M\le |U_N|\le \delta\sqrt N
 \right]\\
 &\le \sum_{p=M}^{[\delta\sqrt{N}]} \Em_{\theta_*}\left[
 |U_N|^k e^{a_\eps U_N^2};
 p\le |U_N|\le p+1
 \right]\\
& \le \sum_{p=M}^{[\delta\sqrt{N}]}(p+1)^k e^{a_\eps(p+1)^2}\Pm_{\theta_*}(|U_N|\ge p).
\enbal
\ee

In order to estimate the probabilities $\Pm_{\theta_*}(|U_N|\ge p)$ in~\eqref{26sep1539}, we make a couple
of preliminary observations.
Since $\theta_*\in\hbox{int}(D)$, there exists $t_0>0$ so that
\be
\Em_{\theta_*}\big[e^{t(X_1-m_*)}\big]<+\infty,~~\hbox{for all $|t|\le t_0$},
\ee
and we can define the log-moment generating function
\be
K(t)=\log\Em_{\theta_*}\big[e^{t(X_1-m_*)}\big],~~|t|\le t_0.
\ee
As we know from~\eqref{26sep1538} that
\[
\Em_{\theta_*}(X_1-m_*)^2=\sigma^2,
\]
we have
\be
K(t)=\farc{\sigma^2t^2}{2}+o(t^2),~~\hbox{as $t\to 0$.}
\ee
{We now fix $b$ such that
\[
a_\eps<b<1/(2\sigma^2).
\]
}
We can decrease $t_0>0$ so that
\be\label{26oct1812}
K(t)\le \farc{t^2}{4b},~~\hbox{for all $|t|\le t_0$.}
\ee
We now fix $\delta>0$ such that, {in addition to~\eqref{26nov1906}, it satisfies}
\[
{0<}\delta\le\min\Big\{h_\eps,\farc{t_0}{2b}\Big\}.
\]
Thus, the parameters $\eps$, $b$, $t_0$, and $\delta$ are fixed independently of $M$ and $N$.
We now use the independence of $X_k$ and~\eqref{26oct1812} to estimate $\Pm_{\theta_*}(|U_N|\ge p)$,
as follows. First, we have for any~$t\in(0,t_0\sqrt{N}]$ and $p\ge 1$,
\be\label{26sep1541}
\bal
\Pm_{\theta_*}(U_N\ge p)&=\Pm_{\theta_*}\big(e^{tU_N}\ge e^{tp}\big)\le e^{-tp}\Em_{\theta_*}(e^{tU_N})=
e^{-tp}\Big(\Em_{\theta_*}\big[e^{t(X_1-m_*)/\sqrt{N}}\big]\Big)^N\\
&=e^{-tp}\exp\Big[NK\Big(\farc{t}{\sqrt{N}}\Big)\Big]\le e^{-tp}e^{Nt^2/(4Nb)}=e^{-tp+t^2/(4b)}.
%e^{-tp}\Big(1+\farc{t^2\sigma^2}{2N}+O\Big(\farc{1}{N^{3/2}}\Big)\Big)^N=
%e^{-tp}e^{t^2\sigma^2/2}+o(1).
\enbal
\ee
Given any~$p\le[\delta\sqrt{N}]$, we know that
\be\label{26nov904}
t_p:=2bp\le t_0\sqrt{N},
\ee
by our choice of $\delta$, which ensures that
\be
\delta\le\farc{t_0}{2b}.
\ee
Thus, we may use~$t=t_p$ in~\eqref{26sep1541}. This gives
%Taking $t=p/\sigma^2$ shows that there exists $b>0$ so that
\be\label{26sep1542}
\Pm_{\theta_*}(U_N\ge p)\le e^{-bp^2},~~\hbox{for all $1\le p\le [\delta\sqrt{N}]$.}
\ee
An identical computation shows that
\be\label{26sep1543}
\Pm_{\theta_*}(U_N\le -p)\le e^{-bp^2},~~\hbox{for all $1 \le p\le [\delta\sqrt{N}]$.}
\ee
Going back to~\eqref{26sep1539} and using~\eqref{26sep1542}-\eqref{26sep1543}, we obtain
\be\label{26sep1544}
\bal
\mathbb E_{\theta_*}\Big[
 |U_N|^k e^{-R_N(U_N)};
 &M\le |U_N|\le \delta\sqrt N
 \Big]
  \le C\sum_{p=M}^{\delta\sqrt{N}}(p+1)^k e^{a_\eps(p+1)^2-bp^2}.
\enbal
\ee
Since $b>a_\eps$ by our choice above, we have
\be\label{26oct1816}
\bal
\mathbb E_{\theta_*}\Big[
 |U_N|^k e^{-R_N(U_N)};
 &~M\le |U_N|\le \delta\sqrt N
 \Big]
  \le C\sum_{p=M}^{\infty}(p+1)^k e^{a_\eps(p+1)^2-bp^2}\\
  &\le C\sum_{p=M}^{\infty}(p+1)^k e^{2bp-(b-a_\eps)p^2}\to 0,~~\hbox{as $M\to+\infty$.}
\enbal
\ee
Now,~\eqref{26oct3010}
follows from~\eqref{26sep1545} and~\eqref{26oct1816}. This
completes the proof of \eqref{26oct2634} and
also finishes the proof of Lemma~\ref{lem-26oct2602}.~$\Box$

\section{The proof of Theorem~\ref{thm:main}}
\label{sec-proof-main-theorem}

In this section, we prove Theorem~\ref{thm:main}. We first use
Lemma~\ref{lem-26oct2602} to obtain the joint Gaussian limit for the
empirical means. Taylor expansions then give the limiting distributions
of the work in part~\textup{(i)}. To deduce parts~\textup{(ii)} and
\textup{(iii)}, we bound the moments of the rescaled work and pass to
the limit in its expectation and variance. The last step treats the
matching and nonmatching cases together.

\subsection{The joint Gaussian limit}
\label{sec-joint-gaussian-limit}

Let us set
\be\label{eq:rescaled-empirical-means}
 U_{i,N}=\sqrt N\,(M_i^{(N)}-m_i),\qquad i=0,1.
\ee
Here, $M_i^{(N)}$ has the Gibbs law $\nu_i^{(N)}$. Thus $U_{i,N}$
describes fluctuations under the Gibbs measure, whereas $U_N$ in
Section~\ref{sec:tilted} describes fluctuations under the exponentially
tilted product measure.

\begin{lemma}\label{lem:joint-gaussian-limit}
The rescaled empirical means satisfy
\begin{equation}\label{eq:work-joint-gaussian-limit}
 (U_{0,N},U_{1,N})
 \Longrightarrow
 \left(\frac{\xi_0}{\sqrt{A_0}},
       \frac{\xi_1}{\sqrt{A_1}}\right),
\end{equation}
where $\xi_0,\xi_1$ are independent standard normal random variables.
\end{lemma}

\begin{proof}
Lemma~\ref{lem-26oct2602} states that, for every integer $k\ge0$,
\be\label{26nov1910}
 \lim_{N\to\infty}\mathbb E[U_{i,N}^{k}]
 =
 \frac{\mathbb E[\xi^{k}e^{-\alpha_i \xi^2}]}
      {\mathbb E[e^{-\alpha_i \xi^2}]},
\ee
where $\xi$ is a centered Gaussian random variable with variance
$\sigma_i^2$.

Since $\sigma_i^{-2}+2\alpha_i=A_i>0$, the right side of~\eqref{26nov1910}
has the form
\be\label{26nov1914}
\begin{aligned}
 \frac{\mathbb E[\xi^{k}e^{-\alpha_i \xi^2}]}
      {\mathbb E[e^{-\alpha_i \xi^2}]}
 &=
 \frac{\displaystyle\int_{\mathbb R}z^k
 e^{-z^2/(2\sigma_i^2)-\alpha_i z^2}\,dz}
 {\displaystyle\int_{\mathbb R}
 e^{-z^2/(2\sigma_i^2)-\alpha_i z^2}\,dz}=
 \int_{\mathbb R}z^k
 \sqrt{\frac{A_i}{2\pi}}\,e^{-A_i z^2/2}\,dz.
\end{aligned}
\ee
Thus all moments of $U_{i,N}$ converge
to those of a  centered Gaussian
with variance $A_i^{-1}$.

{It remains to observe that convergence of moments implies convergence
in distribution. Indeed, Chebyshev's inequality and convergence of the moments shows
that the laws of $U_{i,N}$ are   tight. This is because for any $R>0$ we have
\be
\bal
\Pm\big[|U_{i,N}|>R\big]\le \frac{1}{R^k}
\Em\big[ |U_{i,N}|^k\mathbf 1_{\{|U_{i,N}|>R\}}\big]\le
\frac{1}{R^k}
\Em\big[ |U_{i,N}|^k\big]\le\frac{C_k}{R^k}\to 0,~~\hbox{as $R\to+\infty$,}
\enbal
\ee
uniformly in $N$. Furthermore, any sub-sequential limit of $U_{i,N}$ must have the
Gaussian moments because of~\eqref{26nov1910}-\eqref{26nov1914}. Finally,
the Gaussian distribution is uniquely determined by its moments;
see~\cite[Theorem~30.1 and Example~30.1]{Billingsley}. We deduce that
\[
 U_{i,N}\Longrightarrow \frac{\xi_i}{\sqrt{A_i}}.
\]
Independence of the systems with $i=0,1$ gives the joint convergence
\eqref{eq:work-joint-gaussian-limit}.}
\end{proof}

%shows that the laws of $U_{i,N}$ are {\it tight} in the following sense:
%for every $\varepsilon>0$, a sufficiently large bounded interval
%contains probability at least $1-\varepsilon$, uniformly in $N$.
%Thus every subsequence has a further subsequence converging weakly
%to some probability law.
%
%Let $U$ be a random variable with such a limiting distribution. For every integer $k\ge1$,
%the convergence of the $2k$-th moments gives
%\[
% \sup_N\mathbb E|U_{i,N}|^{2k}<\infty.
%\]
%Consequently,
%\[
% \sup_N\mathbb E\left[
% |U_{i,N}|^k\mathbf 1_{\{|U_{i,N}|>R\}}\right]
% \le \frac{1}{R^k}\sup_N\mathbb E|U_{i,N}|^{2k}
% \longrightarrow0
% \qquad\text{as }R\to\infty.
%\]
%This bound allows us to pass the $k$-th moments to the weak limit.
%Hence $U$ has the Gaussian moments
%\[
% \mathbb E[U^{2k+1}]=0,\qquad
% \mathbb E[U^{2k}]
% =\frac{(2k)!}{2^k k! A_i^k},
% \qquad k\ge0.
%\]
%
%The Gaussian distribution is uniquely determined by its moments;
%see~\cite[Theorem~30.1 and Example~30.1]{Billingsley}.
%Consequently, $U$ has the distribution $\mathcal N(0,A_i^{-1})$.
%Every subsequential limit is therefore this same Gaussian law,
%which proves
%\[
% U_{i,N}\Longrightarrow \frac{\xi_i}{\sqrt{A_i}}.
%\]
%
%Independence gives the joint convergence
%\eqref{eq:work-joint-gaussian-limit}.
%\end{proof}

\subsection{The limiting distribution of the work}
\label{sec-proof-work-distribution}

\begin{proof}[Proof of Theorem~\ref{thm:work-distribution}]
Suppose first that $m_0\ne m_1$. Since \(\Delta H\) is $C^2$,
{there exist $C>0$ and $\ell>0$ so that}
\be\label{26nov1920}
\Delta H(m_i+h)=\Delta H(m_i)+d_i h+r_i(h),
~~ |r_i(h)|\le C h^2,~~
\ee
for \(i=0,1\) and {all \(|h|\le\ell\)}.
{Here, $d_i$ is defined by~\eqref{26oct2728}.}

Recall that
\[
M_i^{(N)}=m_i+\frac{U_{i,N}}{\sqrt N},
\]
{and $\msfw_N$ is given by~\eqref{26oct2318}-\eqref{eq:random-work}:
\begin{equation}\label{26nov1918}
\mathsf w_N=w(M_0^{(N)},M_1^{(N)})=\Delta H(M_0^{(N)})-\Delta H(M_1^{(N)}).
\end{equation}
}

Substituting this into the two Taylor expansions~\eqref{26nov1920}
and subtracting, we obtain
\be\label{26nov1923}
\sqrt N\,(\mathsf w_N-w_\infty)
=d_0U_{0,N}-d_1U_{1,N}+\mathcal R_N,
\ee
where
\[
\mathcal R_N=\sqrt N\left[
r_0\!\left(\frac{U_{0,N}}{\sqrt N}\right)
-r_1\!\left(\frac{U_{1,N}}{\sqrt N}\right)
\right].
\]

We now show that \(\mathcal R_N\) converges to zero in probability. By tightness, for every \(\varepsilon>0\) we can choose $K_\eps>0$ so that
\be\label{26nov1921}
\mathbb P\bigl(|U_{0,N}|\le K_\eps,\ |U_{1,N}|\le K_\eps\bigr)
\ge 1-\varepsilon
\ee
for all \(N\). On this event, for sufficiently large \(N\), {we have
\be
\frac{|U_{i,N}|}{\sqrt{N}}<\ell,
\ee
and~\eqref{26nov1920} implies}
\be\label{26nov1922}
|\mathcal R_N|
\le \frac{C}{\sqrt N}
\bigl(U_{0,N}^2+U_{1,N}^2\bigr)
\le \frac{2CK_\eps^2}{\sqrt N}.
\ee
{Combining~\eqref{26nov1921} and~\eqref{26nov1922}, we see that
\be\label{26nov1924}
\hbox{$\mathcal R_N\to 0$ in probability,}
\ee
as claimed.}
By~\eqref{eq:work-joint-gaussian-limit} and~\eqref{26nov1924},
the right side of~\eqref{26nov1923}
converges in distribution to
\[
 \frac{d_0}{\sqrt{A_0}}\xi_0
 -\frac{d_1}{\sqrt{A_1}}\xi_1,
\]
which is a centered Gaussian with variance $V$.
This proves~\eqref{eq:work-distribution-nonmatching}.

Now suppose that $m_0=m_1=m_*$. {Since $m_*$ is the minimum of both $\Phi_{0}$ and $\Phi_1$, we have}
\[
 I'(m_*)+\beta H_i'(m_*)=0,\qquad i=0,1.
\]
{Subtracting these equations, we see that}
\be\label{26nov1926}
\Delta H'(m_*)=0.
\ee
Moreover,
{recalling the definitions~\eqref{26oct2912} of~$A_i$ and~\eqref{26nov1904} of~$c$, we obtain}
\[
 A_1-A_0
 =\beta\bigl(H_1''(m_*)-H_0''(m_*)\bigr)
 =\beta c.
\]
The second-order Taylor expansion {near $m_*$} has the form
\be\label{26sep1421}
\Delta H(m_*+h)
 =
 \Delta H(m_*)
 +
 \frac{\Delta H''(m_*)}{2}h^2
 +
 h^2q(h),
 \qquad
 \lim_{h\to 0}q(h) = 0.
\ee
Therefore, in this case we have
\[
 N\mathsf w_N
 =\frac c2\bigl(U_{0,N}^2-U_{1,N}^2\bigr)
   +o_{\mathbb P}(1).
\]
{The last term above converges to zero in probability by essentially the
same argument as in the proof of~\eqref{26nov1924}.}
\end{proof}
%Indeed, on the event $|U_{0,N}|\le K$, $|U_{1,N}|\le K$, the absolute
%value of the remainder is at most
%\[
% 2K^2\sup_{|h|\le K/\sqrt N}|q(h)|,
%\]
%where we set $q(0)=0$. For each fixed $K$, this expression tends to
%zero as $N\to+\infty$. Tightness, as in the preceding argument,
%therefore shows that the remainder tends to zero in probability.
%The continuous mapping theorem and
%\eqref{eq:work-joint-gaussian-limit} now prove
%\eqref{eq:work-distribution-matching}.

%Finally, for a standard normal variable $\xi$, we have
%\[
% \mathbb E[e^{iu\xi^2}]=(1-2iu)^{-1/2}.
%\]
%Applying this identity to the two independent terms defining $\zeta$
%gives~\eqref{eq:work-distribution-characteristic}.

%\end{proof}

\subsection{Bounds on the moments of the rescaled work}
\label{sec-work-moment-bounds}

Let us set
\be\label{eq:work-rescaling}
 a_N=
 \begin{cases}
 \sqrt N,&m_0\ne m_1,\\
 N,&m_0=m_1,
 \end{cases}
 \qquad
 Y_N=a_N(\mathsf w_N-w_\infty).
\ee
Recall that $w_\infty=0$ when $m_0=m_1$. Thus $Y_N$ is precisely
the rescaled random variable appearing in
\eqref{eq:work-distribution-nonmatching} or
\eqref{eq:work-distribution-matching}, respectively.

The next lemma allows us to pass from convergence in distribution to
convergence of expectations and variances. For this purpose, a bound
on the fourth moment is sufficient. We state the bound for all even
moments, since the proof is the same and also yields the convergence
of all moments asserted after Theorem~\ref{thm:main}.

\begin{lemma}[Bounds on the moments]\label{lem:work-moment-bounds}
For every integer $k\ge1$, there exist $C_k>0$ and $N_k\in\N$ such
that
\be\label{eq:work-even-moment-bound}
 \mathbb E|Y_N|^{2k}\le C_k,\qquad N\ge N_k.
\ee
In particular, the fourth moments of $Y_N$ are bounded for all
sufficiently large $N$.
\end{lemma}

\begin{proof}
By the definitions of $\mathsf w_N$ and $w_\infty$, we have
\[
 Y_N=a_N\big[\Delta H(M_0^{(N)})-\Delta H(m_0)
              -\Delta H(M_1^{(N)})+\Delta H(m_1)\big].
\]
Using the elementary inequality
$|x-y|^{2k}\le 2^{2k-1}(|x|^{2k}+|y|^{2k})$, we obtain
\be\label{eq:work-moment-reduction}
 \mathbb E|Y_N|^{2k}
 \le 2^{2k-1}\sum_{i=0}^1 a_N^{2k}
 \int_{\mathbb R}|\Delta H(m)-\Delta H(m_i)|^{2k}
                  \,d\nu_i^{(N)}(m).
\ee
It remains to bound the two terms on the right, uniformly for all
sufficiently large $N$.

Fix $i=0,1$ and a sufficiently small $\delta>0$. We split the
integral in~\eqref{eq:work-moment-reduction} into the regions
$|m-m_i|\le\delta$ and $|m-m_i|>\delta$.
By the assumption~\eqref{26oct2602}, there exist constants
$C,\kappa>0$, which may depend on $k$ but not on $N$, such that
\[
 |\Delta H(m)-\Delta H(m_i)|^{2k}\le C e^{\kappa|m|}.
\]
The concentration estimate~\eqref{26oct2723} in
Lemma~\ref{lem:26oct2704}, used with $r=\delta$, now implies
\be\label{eq:rescaled-work-tail}
 \bal
 &a_N^{2k}\int_{|m-m_i|>\delta}
      |\Delta H(m)-\Delta H(m_i)|^{2k}\,d\nu_i^{(N)}(m)\\
 &\qquad\le C a_N^{2k}\int_{|m-m_i|>\delta}
             e^{\kappa|m|}\,d\nu_i^{(N)}(m)
 \le C N^{2k}e^{-\gamma N},
 \enbal
\ee
for some $C,\gamma>0$ and all sufficiently large $N$. Here we used
$a_N\le N$. In particular, this {term}  tends to zero as $N\to+\infty$.

At the same time,   we can use the
Taylor expansion of $\Delta H$ in the region $|m-m_i|\le\delta$. If $m_0\ne m_1$, then
\[
 |\Delta H(m)-\Delta H(m_i)|\le C|m-m_i|,
 \qquad |m-m_i|\le\delta,
\]
and $a_N=\sqrt N$. Therefore,
\be\label{eq:work-local-moment-nonmatching}
 \bal
 &a_N^{2k}\int_{|m-m_i|\le\delta}
      |\Delta H(m)-\Delta H(m_i)|^{2k}\,d\nu_i^{(N)}(m)\\
 &\qquad\le C N^k\int_{\mathbb R}|m-m_i|^{2k}\,d\nu_i^{(N)}(m)
 =C\mathbb E|U_{i,N}|^{2k}.
 \enbal
\ee
The last expression is bounded by Lemma~\ref{lem-26oct2602}.

If $m_0=m_1=m_*$, then $\Delta H'(m_*)=0$, by~\eqref{26nov1926}.
Choose~$\delta>0$ so that the Taylor
expansion~\eqref{26sep1421} holds for $|h|\le\delta$ and $q(h)$
is bounded on this interval. It follows that
\[
 |\Delta H(m)-\Delta H(m_*)|\le C|m-m_*|^2,
 \qquad |m-m_*|\le\delta.
\]
Since $a_N=N$, we obtain
\be\label{eq:work-local-moment-matching}
 \bal
 &a_N^{2k}\int_{|m-m_*|\le\delta}
      |\Delta H(m)-\Delta H(m_*)|^{2k}\,d\nu_i^{(N)}(m)\\
 &\qquad\le C N^{2k}\int_{\mathbb R}|m-m_*|^{4k}\,d\nu_i^{(N)}(m)
 =C\mathbb E|U_{i,N}|^{4k}.
 \enbal
\ee
This expression is also bounded by Lemma~\ref{lem-26oct2602}.
Combining~\eqref{eq:rescaled-work-tail} with
\eqref{eq:work-local-moment-nonmatching} or
\eqref{eq:work-local-moment-matching}, and then
using~\eqref{eq:work-moment-reduction}, proves
\eqref{eq:work-even-moment-bound}.
\end{proof}

\subsection{The expectations and standard deviations}
\label{sec-proof-work-moments}

We now deduce parts~\textup{(ii)} and~\textup{(iii)} of
Theorem~\ref{thm:main} simultaneously for the matching and
nonmatching cases.

According to part~\textup{(i)} proved above, $Y_N$ converges in distribution to a random variable~$Y$.
In the non-matching case $m_0\ne m_1$, the limiting law is
$\mathcal N(0,V)$: thus $Y$ is a centered Gaussian random variable
with variance $V$. In the matching case $m_0=m_1$, we take
\[
 Y=\zeta=\frac{c}{2A_0}\xi_0^2-\frac{c}{2A_1}\xi_1^2,
\]
where $\xi_0,\xi_1$ are independent standard normal random variables,
each with mean zero and variance one.

We claim that
\be\label{26nov1927}
 \mathbb EY_N\longrightarrow\mathbb EY,
~~
 \mathbb EY_N^2\longrightarrow\mathbb EY^2.
\ee
To prove this claim, we note that the case $k=2$ in
Lemma~\ref{lem:work-moment-bounds} {implies a uniform bound
on the fourth moment of $Y_N$:} there exist $C>0$ and
$N_0\in\N$ such that
\be
\mathbb E|Y_N|^4\le C,~~\hbox{ for all $N\ge N_0$.}
\ee
{This gives the uniform integrability of the first two moments:}
\be\label{eq:work-uniform-integrability}
 \sup_{N\ge N_0}
 \mathbb E\big[|Y_N|^j\mathbf1_{\{|Y_N|>R\}}\big]
 \le \frac{C}{R^{4-j}}\to0,\qquad\hbox{as $R\to+\infty$,~{for $j=1,2$.}}
\ee
Together, uniform integrability of the first second moment
and convergence in probability imply~\eqref{26nov1927}.

It remains only to {compute}
the moments of the limiting distributions.
For the Gaussian limit they are
\[
 \mathbb EY=0,\qquad \operatorname{Var}(Y)=V.
\]
For the matching limit, using the independence of $\xi_0,\xi_1$,
$\mathbb E\xi_i^2=1$, and $\operatorname{Var}(\xi_i^2)=2$, we obtain
\be\label{eq:matching-limit-moments}
 \mathbb E\zeta=\frac c2\left(\frac1{A_0}-\frac1{A_1}\right),
 \qquad
 \operatorname{Var}(\zeta)
 =\frac{c^2}{2}\left(\frac1{A_0^2}+\frac1{A_1^2}\right).
\ee
Since $c=(A_1-A_0)/\beta$, the first expression equals
\[
 \mathbb E\zeta=\frac{(A_1-A_0)^2}{2\beta A_0A_1}.
\]
These formulas give~\eqref{26oct2716},
\eqref{26oct2718}, \eqref{26oct2730}, and~\eqref{26oct2731}.
The strict positivity of $w_\infty$ when $m_0\ne m_1$ was explained
after Theorem~\ref{thm:main}. This finishes the proof of the theorem.
\Qed

\begin{rem} \label{rem-allmoments} {\rm
The same argument, {together with Lemma~\ref{lem:work-moment-bounds}}
gives convergence of all moments.
%Indeed, fix an
%integer $j\ge1$ and choose $k$ so that $2k>j$. By
%Lemma~\ref{lem:work-moment-bounds},
%\[
% \sup_{N\ge N_k}
% \mathbb E\big[|Y_N|^j\mathbf1_{\{|Y_N|>R\}}\big]
% \le \frac{C_k}{R^{2k-j}}\to0,\qquad\hbox{as $R\to+\infty$.}
%\]
%Applying convergence in distribution to the bounded continuous
%function $g_R(x)^j$ and then letting $R\to+\infty$ gives
%$\mathbb EY_N^j\to\mathbb EY^j$.
This proves the additional
assertion about moments following Theorem~\ref{thm:main}.
}
\end{rem}

\section{An example: Gaussian oscillators}
\label{sec:gaussian-example}

In this section, we consider Gaussian oscillators, an example of the
general systems we have discussed in the previous sections. The advantage
of this simple system is that one can compute exactly the Gibbs measures,
the distribution of the work, its expectation and its variance to compare
with the general results of Theorem~\ref{thm:main}. We discuss these
three aspects of the work in the same order as in that theorem.

 \subsection{The Gibbs measure}
\label{subsec-Gibbs-measure-Gaussian-oscillators}

Consider $N$ independent oscillators with displacements $s_i$ distributed
according to the normal distribution $\mathcal N(0,1)$.  We assume that the energy per particle
has the form
\begin{equation} \label{26oct2814}
 H_{q,b}(m)=bm^2+qm,
\end{equation}
with the mean displacement $m$ given by~\eqref{26oct2812}, the coupling constant $b>0$
and $q\in\Rm$. The microscopic Gaussian density is
\be
\rho(s)=\farc{1}{\sqrt{2\pi}}e^{-|s|^2/2},
\ee
%We fix $\beta>0$ and assume $b>0$.  The $N$-particle energy is
%$NH_{q,b}(m)$.
and  the Gibbs measure on $\mathbb R^N$ is
\[
d\mu_{q,b}^{(N)}(s_1,\dots,s_N)
=Z_N(q,b)^{-1}
e^{-\beta NH_{q,b}(m)}\,d\gamma_N(s_1,\dots,s_N),
\]
with
\[
 d\gamma_N
 =
 (2\pi)^{-N/2}
 \exp\Big\{-\frac12\sum_{i=1}^Ns_i^2\Big\}
 ds_1\cdots ds_N.
\]

 \subsubsection{The partition function}

The partition function
\[
Z_N =\int_{\Rm^N} e^{-\frac12\sum_i s_{i}^2}
e^{-\beta N(bm^2+qm)}\,\farc{ds_1\dots ds_N}{(2\pi)^{N/2}}
\]
can be computed as follows. Introduce the orthonormal basis
\be
\bal
&v_k=\farc{1}{\sqrt{k+k^2}}(e_1+\dots+e_k-ke_{k+1}),~~k=1,\dots,N-1,\\
&v_N=\farc{1}{\sqrt{N}}(e_1+\dots+e_N).
\enbal
\ee
The corresponding coordinates are
\be\label{26oct2820}
\bal
&u_{k}=(s\cdot v_k)=\farc{\sqrt{k}}{\sqrt{k+1}}\Big(\farc{s_1+\dots+s_{k}}{k}-s_{k+1}\Big),~~k=1,\dots,N-1,\\
&u_N=(s\cdot v_N)=\farc{s_1+\dots+s_N}{\sqrt{N}}.
\enbal
\ee
Then we have
\be
\sum_{k=1}^Nu_k^2=\sum_{k=1}^N s_k^2,~~m=\farc{1}{\sqrt{N}}u_N,
\ee
and
\be\label{26oct2821}
\bal
Z_N &=\int_{\mathbb R^N} e^{-\frac12\sum_i s_i^2} e^{-\beta N(bm^2+qm)}\,
\farc{ds_1\dots ds_N}{(2\pi)^{N/2}}
=\int_{\mathbb R^N} e^{-\frac12\sum_i u_i^2} e^{-\beta (bu_N^2+\sqrt{N}qu_N)}\,
\farc{du_1\dots du_N}{(2\pi)^{N/2}}.
\enbal
\ee
Integrating out $u_1,\dots,u_{N-1}$ and setting $x=\sqrt{N}u_N$ gives
%Introduce
%\[
% x = \sum_{i=1}^N s_i,
% \quad m=\frac{x}{N},
%\]
%and extend to orthonormal coordinates $(u_1,\dots,u_N)$ by
%\[
% u_1 = \frac{x}{\sqrt N},
% \quad u_j\ (j\ge2)=\text{orthogonal modes},
%\]
%so that $\sum_i s_i^2 = u_1^2+\sum_{j=2}^N u_j^2$ and the Jacobian is $d^Ns = du_1\cdots du_N$.  Integrating out $u_2,\dots,u_N$ gives a factor $(2\pi)^{(N-1)/2}$, yielding
%\[
% Z_N = (2\pi)^{-1/2}\int_{\mathbb R} e^{-\frac12 u_1^2} e^{-\beta N(bu_1^2/N+qu_1/\sqrt N)} du_1.
%\]
%Set $x=\sqrt N\,u_1$ (so $du_1=dx/\sqrt N$) to get
\be\label{26oct2826}
\bal
Z_N &= \int_{\mathbb R} \exp\Bigl[-\farc{x^2}{2N}-\beta\Bigl(b\farc{x^2}{N}+ q x\Bigr)\Bigr]
\farc{dx}{\sqrt{2\pi N}}
=\int_{\mathbb R} \exp\Bigl[-\frac{(1+2b\beta)x^2}{2N}-\beta q x\Bigr]\farc{dx}{\sqrt{2\pi N}} \\
&=
\int_{\mathbb R} \exp\Bigl[-\frac{y^2}{2N}-\farc{\beta q}{\sqrt{1+2b\beta}}y\Bigr]
\farc{dy}{\sqrt{2\pi (1+2b\beta)N}}\\
&=\int_{\mathbb R}
\exp\Bigl[-\farc{1}{2N}\Big(y+\farc{N\beta q}{\sqrt{1+2b\beta}}\Big)^2+\farc{N\beta^2 q^2}
{2(1+2b\beta)}\Bigr]\farc{dy}{\sqrt{2\pi (1+2b\beta)N}}\\
&=\farc{1}{\sqrt{ (1+2b\beta)}}
\exp\Bigl[\farc{N\beta^2 q^2}{2(1+2b\beta)}\Bigr].
\enbal
\ee
 Thus, for fixed $q$, at  equilibrium,  the free energy per oscillator is
\begin{equation}\label{26oct2902}
-\frac{1}{\beta N}\ln Z_N = -\frac{\beta q^2}{2(1+2\beta b)} + O\bigl(N^{-1}\bigr).
\end{equation}

 \subsubsection{The Gibbs measure on $m$}

Next, we explicitly
compute the push-forward $\nu_{q,b}^{(N)}(m)$ of the measure $d\mu_{q,b}^{(N)}$ to the
macroscopic variable $m$.
 Take any continuous compactly supported function $f(m)$. A computation
very close to that in~\eqref{26oct2820}-\eqref{26oct2821} shows that
\be
\bal
\int_{\Rm^N}f(m) d\mu_{q,b}^{(N)}&= Z_N^{-1}\int_{\mathbb R} f\Big(\farc{x}{N}\Big)
\exp\Bigl[-\farc{x^2}{2N}-\beta\Bigl(b\farc{x^2}{N}+ q x\Bigr)\Bigr]
\farc{dx}{\sqrt{2\pi N}}
\\
&=Z_N^{-1}\int_{\mathbb R} f(y)
\exp\Bigl[-\frac{(1+2b\beta)Ny^2}{2}-\beta q Ny\Bigr]\farc{\sqrt{N}dy}{\sqrt{2\pi }}.
\enbal
\ee
We may now use~\eqref{26oct2826} to obtain
\be\label{26oct2904}
\bal
\int_{\Rm^N}f(m) d\mu_{q,b}^{(N)}
&= \int_{\Rm} f(y)\exp\Bigl[-\farc{(1+2b\beta) N}{2}\Big[y^2+
\farc{2\beta q y}{1+2b\beta}+\farc{\beta^2q^2}{(1+2b\beta )^2}\Bigr]
\farc{\sqrt{(1+2b\beta)N}dy}{\sqrt{2\pi }}\\
&=
\int_{\mathbb R}  f(y)\exp\Bigl[-\farc{(1+2b\beta)N}{2}
\Big(y+\farc{\beta q}{1+2b\beta}\Big)^2 \Bigr]
\farc{\sqrt{(1+2b\beta)N}dy}{\sqrt{2\pi  }}.
\enbal
\ee
We see from~\eqref{26oct2904} that the Gibbs measure is a Gaussian
\be\label{26oct2906}
d\nu_{q,b}^{(N)}(m)=\farc{1}{\sqrt{2\pi\tau_N^2(q,b)}}
e^{-|m-m_*(q,b)|^2/(2\tau_N^2(q,b))} dm\,
\ee
with the mean
\be\label{26oct2908}
m_*(q,b)=-\frac{\beta q}{1+2\beta b},
\ee
and the variance
\be\label{26nov1929}
\tau_N^2(q,b)=\farc{1}{(1+2b\beta)N}\to 0,~~\hbox{as $N\to+\infty$.}
\ee
In particular, we have
a weak convergence
\[
\nu_{q,b}^{(N)}\Rightarrow \delta(m-m_*(q,b)),~~\hbox{as $N\to+\infty$.}
\]

\subsection{Work over a jump cycle}

In this section we discuss what happens when the system described in
Section~\ref{subsec-Gibbs-measure-Gaussian-oscillators} undergoes an
instantaneous jump.
Suppose that the stiffness parameter changes instantaneously from $b_0$ to $b_1$ and the external
field changes from $q_0$ to $q_1=q_0+\delta q$.  When $b_1\neq b_0$, the two
limiting equilibrium macrostates $m_*(q_0,b_0)$ and $m_*(q_1,b_1)$ given by~\eqref{26oct2908}
coincide if
\begin{equation}\label{26oct2916}
 q_0=q_0^*:=\delta q\,\frac{1+2\beta b_0}{2\beta(b_1-b_0)}.
\end{equation}
For this value, the transition was identified in
\cite{EPR-JSP,EPR-LMP} as ultrafast in the thermodynamic limit.

If the parameters are then changed back from $(q_0^*+\delta q,b_1)$ to
$(q_0^*,b_0)$, the reverse transition is ultrafast as well.  \color{black}
At the level of the limiting equilibrium macrostates, this may look like a reversible cycle driven by instantaneous changes of the Hamiltonian, whereas reversible processes in standard thermodynamics are usually quasistatic; see, for example,
\cite[Section~4-3]{Callen}. In fact this apparent reversibility motivated our study of the work required to perform the cycle. \color{black} Proposition~\ref{prop:work-positive} and
Theorem~\ref{thm:work-asympt} show that the finite-$N$ cycle is nevertheless
irreversible whenever the two finite-$N$ Gibbs measures differ. Next, we compute
the distribution of the work, its expectation and its variance for the
Gaussian oscillators.

Let us set
\[
\begin{aligned}
&H_i(m)=H_{q_i,b_i}(m),\qquad
 \nu_i^{(N)}=\nu_{q_i,b_i}^{(N)},\\
&m_i=m_*(q_i,b_i),\qquad
 \tau_{i,N}=\tau_N(q_i,b_i),\qquad i=0,1.
\end{aligned}
\]
Since $I(m)=m^2/2$ for the standard normal density, we have
\[
 \Phi_i(m)=\frac{m^2}{2}+\beta(b_i m^2+q_i m),
 \qquad A_i=\Phi_i''(m_i)=1+2\beta b_i.
\]
We also write
\[
 B=b_1-b_0,\qquad
 \Delta H(m)=Bm^2+(q_1-q_0)m,
\]
and, in accordance with the notation of Section~\ref{sec:work}, set
\[
 d_i=\Delta H'(m_i)=2Bm_i+q_1-q_0,\qquad
 w_\infty=\Delta H(m_0)-\Delta H(m_1).
\]

\subsubsection{Distribution of the work}

Recall that the empirical means $M_0^{(N)}$ and $M_1^{(N)}$ at the
two jumps are independent and have the laws $\nu_0^{(N)}$ and
$\nu_1^{(N)}$, respectively. We see from~\eqref{26oct2906} that
\begin{equation}\label{eq:gaussian-empirical-means}
 (M_0^{(N)},M_1^{(N)})\overset{\mathrm{law}}=
 \left(m_0+\frac{\xi_0}{\sqrt{NA_0}},
       m_1+\frac{\xi_1}{\sqrt{NA_1}}\right),
\end{equation}
where $\xi_0,\xi_1$ are independent standard normal random variables.
Since $\Delta H$ is quadratic, the identity
\[
 \Delta H(m_i+u)=\Delta H(m_i)+d_i u+Bu^2
\]
holds for every $u\in\Rm$. Substituting~\eqref{eq:gaussian-empirical-means}
into the expression~\eqref{eq:random-work} for the random work per
particle gives
\begin{equation}\label{eq:gaussian-random-work}
\begin{aligned}
 \mathsf w_N\overset{\mathrm{law}}={}&w_\infty
 +\frac1{\sqrt N}
  \left(\frac{d_0}{\sqrt{A_0}}\xi_0
       -\frac{d_1}{\sqrt{A_1}}\xi_1\right)+\frac{B}{N}
  \left(\frac{\xi_0^2}{A_0}-\frac{\xi_1^2}{A_1}\right).
\end{aligned}
\end{equation}
This formula is exact for every $N\ge1$.

Suppose first that $m_0\ne m_1$. After subtracting $w_\infty$ and
multiplying~\eqref{eq:gaussian-random-work} by $\sqrt N$, the quadratic
term tends to zero in probability. Hence
\begin{equation}\label{eq:gaussian-work-nonmatching}
 \sqrt N\,(\mathsf w_N-w_\infty)
 \Longrightarrow\mathcal N(0,V),\qquad
 V=\frac{d_0^2}{A_0}+\frac{d_1^2}{A_1},
\end{equation}
as in Theorem~\ref{thm:work-distribution}. In this example the limiting
Gaussian is always non-degenerate when $m_0\ne m_1$. Indeed,
using~\eqref{26oct2908} to write $q_i=-A_i m_i/\beta$, we obtain
\begin{equation}\label{eq:gaussian-work-slopes}
 d_0=\frac{A_1}{\beta}(m_0-m_1),\qquad
 d_1=\frac{A_0}{\beta}(m_0-m_1).
\end{equation}
Thus $V>0$ in the nonmatching case.

Suppose now that $m_0=m_1=m_*$. Then $w_\infty=d_0=d_1=0$, and
\eqref{eq:gaussian-random-work} reduces to
\begin{equation}\label{eq:gaussian-work-matching}
 N\mathsf w_N\overset{\mathrm{law}}=
 \frac{B}{A_0}\xi_0^2-\frac{B}{A_1}\xi_1^2,
 \qquad N\ge1.
\end{equation}
Since $\Delta H''(m_*)=2B=(A_1-A_0)/\beta$, the random variable on
the right is precisely the matching limit $\zeta$ in
Theorem~\ref{thm:work-distribution}. Thus, for Gaussian oscillators,
the distribution of the total work at matching is independent of $N$:
the limiting law is already exact for every finite $N$.
Its characteristic function is
\begin{equation}\label{eq:gaussian-work-characteristic}
 \mathbb E[e^{itN\mathsf w_N}]
 =\left(1-\frac{2itB}{A_0}\right)^{-1/2}
  \left(1+\frac{2itB}{A_1}\right)^{-1/2},\qquad t\in\Rm,
\end{equation}
where the square roots are chosen continuously with value $1$ at $t=0$.
When $b_0\ne b_1$, this law is non-Gaussian and non-degenerate.
Although the fluctuations of the empirical means are Gaussian, the
linear terms in the work vanish at matching, and the work is determined
by the squares of these fluctuations. If $b_0=b_1$ and $m_0=m_1$,
then $q_0=q_1$ as well, and the work vanishes identically.

\subsubsection{Expectation of the work}

We now compute the expectation of the random work in~\eqref{eq:gaussian-random-work}.
The average work per particle over the cycle is
\begin{equation}\label{26oct2918}
\bal
 w_N(q_0,b_0;q_1,b_1)&=
 \int_{\mathbb R}(H_1(m)-H_0(m))\,d\nu_0^{(N)}
 -
 \int_{\mathbb R}(H_1(m)-H_0(m))\,d\nu_1^{(N)}\\
 &=\int_{\mathbb R}((b_1-b_0)m^2+(q_1-q_0)m)\,d\nu_0^{(N)}(m)
 \\
 &-\int_{\mathbb R}((b_1-b_0)m^2+(q_1-q_0)m)\,d\nu_1^{(N)}.
\enbal
\end{equation}
We see from the explicit expression~\eqref{26oct2906} that
\be\label{26oct2920}
 \int_{\mathbb R}m\,d\nu_i^{(N)}=m_i,
 \qquad
 \int_{\mathbb R}m^2\,d\nu_i^{(N)}
 =
 m_i^2+\tau_{i,N}^2.
\ee
Using these expressions in~\eqref{26oct2918} gives
\be\label{26oct2923}
\bal
w_N(q_0,b_0;q_1,b_1)&=(b_1-b_0)\big(m_0^2+\tau_{0,N}^2\big)+
(q_1-q_0)m_0-(b_1-b_0)\big(m_1^2+\tau_{1,N}^2\big)-(q_1-q_0)m_1\\
&=(m_0-m_1)\big((b_1-b_0)(m_0+m_1)+q_1-q_0\big)
+(b_1-b_0)(\tau_{0,N}^2-\tau_{1,N}^2).
\enbal
\ee
Let us use~\eqref{26oct2908} to replace $q_{0,1}$ in the first term above:
\be\label{26oct2926}
\bal
(b_1-b_0)&(m_0+m_1)+q_1-q_0=(b_1-b_0)(m_0+m_1)-\farc{1+2\beta b_1}{\beta}m_1+
\farc{1+2\beta b_0}{\beta}m_0\\
&=(b_1-b_0)(m_0+m_1)-2b_1m_1+2b_0m_0+\farc{1}{\beta}(m_0-m_1)\\
&=(b_1+b_0)(m_0-m_1)+\farc{1}{\beta}(m_0-m_1)=\big(b_1+b_0+\beta^{-1}\big)(m_0-m_1).
\enbal
\ee
The last term in the right side of~\eqref{26oct2923} can be written as
\be\label{26oct2925}
\tau_{0,N}^2-\tau_{1,N}^2=\farc{1}{(1+2b_0\beta)N}-\farc{1}{(1+2b_1\beta)N}=
\farc{2\beta(b_1-b_0)}{(1+2b_0\beta)(1+2b_1\beta)N}.
\ee
Using~\eqref{26oct2926} and~\eqref{26oct2925} in~\eqref{26oct2923}, we obtain
\be
\bal
w_N(q_0,b_0;q_1,b_1)&=\big(b_1+b_0+\beta^{-1}\big)(m_0-m_1)^2+
\farc{2\beta(b_1-b_0)^2}{(1+2b_0\beta)(1+2b_1\beta)N}.
\enbal
\ee
Both terms in the right side above are non-negative and their sum
vanishes only if $b_0=b_1$ and $m_0=m_1$, which means that $q_1=q_0$ as well.
As in Theorem~\ref{thm:work-asympt},
we see that if $m_0\neq m_1$ then the average work per particle $w_N$ remains positive in the limit
$N\to+\infty$. On the other hand, in the matching case $m_0=m_1$ we have
$w_N=O(1/N)$ in the thermodynamic limit.

\subsubsection{Variance and standard deviation of the work}

Let us return to~\eqref{eq:gaussian-random-work}. Since
\[
 \mathbb E[\xi_i]=\mathbb E[\xi_i^3]=0,\qquad
 \mathbb E[\xi_i^2]=1,\qquad
 \mathbb E[\xi_i^4]=3,
\]
we have
\[
 \operatorname{Cov}(\xi_i,\xi_i^2)=0,\qquad
 \operatorname{Var}(\xi_i^2)=2.
\]
Together with the independence of $\xi_0$ and $\xi_1$, this shows that
the covariance between the linear and quadratic terms in
\eqref{eq:gaussian-random-work} vanishes. Consequently, the variance of
the work per particle is
\begin{equation}\label{eq:gaussian-work-variance}
 \sigma_N^2=\operatorname{Var}(\mathsf w_N)
 =\frac1N\left(\frac{d_0^2}{A_0}+\frac{d_1^2}{A_1}\right)
  +\frac{2B^2}{N^2}\left(\frac1{A_0^2}+\frac1{A_1^2}\right).
\end{equation}
Here $\sigma_N$ denotes the standard deviation of the work, as
in~\eqref{eq:work-standard-deviation}.
The quantities $\tau_{i,N}$ that appear in~\eqref{26nov1929}
are the standard deviations of the empirical means.

If $m_0\ne m_1$, we obtain
\[
 \sqrt N\,\sigma_N\longrightarrow
 \left(\frac{d_0^2}{A_0}+\frac{d_1^2}{A_1}\right)^{1/2}>0.
\]
In the matching case $m_0=m_1$, both $d_0$ and $d_1$ vanish, and
\eqref{eq:gaussian-work-variance} gives the exact formula
\begin{equation}\label{eq:gaussian-work-matching-deviation}
 N\sigma_N=\sqrt2\,|B|
 \left(\frac1{A_0^2}+\frac1{A_1^2}\right)^{1/2}.
\end{equation}
Since $\Delta H''(m_*)=2B$, these formulas agree with
Theorem~\ref{thm:work-fluct}. At matching, both the mean and the
standard deviation of the total work are independent of $N$.

As explained in  \cite{EPR-JSP,EPR-LMP} and also in Section~\ref{sec:reeb} above,
the matching condition $m_0=m_1$ corresponds to
a Reeb chord connecting the equilibrium Legendrian submanifolds. Interestingly,
for Gaussian particles and a quadratic Hamiltonian, this
interpretation holds for every finite $N$ and does not need concentration
to a delta measure.

\appendix

\section{Basic large deviation results}
\label{app:large-dev}

We recall the two standard large deviation results used in the proof of
Theorem~\ref{thm:main}. We refer to \cite{DZ} for details.

Let $X_1,X_2,\ldots$ be independent identically distributed real-valued random variables, and let
\[
 \Lambda(\theta)
 =
 \log\mathbb E \big[e^{\theta X_1}\big],
~~
 D=\{ \theta:\Lambda( \theta)<+\infty\}.
\]
Assume that $0\in\operatorname{int}D$, and let $P_N$ be the distribution of
\[
m_N=\frac1N\sum_{j=1}^NX_j.
\]
Define
\[
 I( m)
 =
 \sup_{\theta\in\mathbb R}
 \left\{
\theta m-\Lambda(\theta)
 \right\}.
\]

\begin{thm}[Cram\'er's theorem]\label{thm:cramer}
The sequence $P_N$ satisfies a large deviation principle with speed $N$
and good\footnote{A rate function is called
\emph{good} if all its sublevel sets
$\{m\in\mathbb R:I(m)\leq a\}$, with $a<\infty$, are compact.} convex rate function $I$.
That is, for every closed set
$F\subset\mathbb R$ and every open set $G\subset\mathbb R$,
\be\label{26oct2614}
 \limsup_{N\to +\infty}\frac1N\log P_N(F)
 \leq-\inf_F I,
\ee
and
\be\label{26oct2616}
 \liminf_{N\to +\infty}\frac1N\log P_N(G)
 \geq-\inf_G I.
\ee
\end{thm}

\begin{thm}[Varadhan's lemma]\label{thm-lem-varadhan}
Let $F:\mathbb R\to\mathbb R$ be continuous and bounded from above.
Then
\[
 \lim_{N\to +\infty}
 \frac1N\log
 \int_{\mathbb R}e^{NF(  m)}\,P_N(d  m)
 =
 \sup_{  m\in\mathbb R }
 \{F(  m)-I( m)\}.
\]
\end{thm}
We also use the standard exponential-tilting consequence.  If
\[
 Z_N=\int_{\mathbb R}e^{NF(m)}\,P_N(dm),
 \qquad
 Q_N(dm)=Z_N^{-1}e^{NF(m)}P_N(dm),
\]
then $Q_N$ satisfies a large deviation principle with rate function
\begin{equation}\label{26oct2930}
J(m)=I(m)-F(m)-
 \inf_{m'}
 \bigl(I(m')-F(m')\bigr).
\end{equation}

 \section{The log-moment generating function and the rate function}
\label{app:log-mgf-rate}

\begin{proof}[Proof of Proposition~\ref{prop:log-mgf-rate}]
\emph{Smoothness and strict convexity of~$\Lambda$.}
Since~$D$ is open, differentiation under the integral in~\eqref{26oct2932}
is justified to every order, locally uniformly in~$\theta\in D$.
Indeed, exponential moments at two nearby parameters dominate the
integrands and all their derivatives on any smaller parameter interval.
Thus~$\Lambda$ is $C^\infty$ on~$D$.
For $\theta\in D$, introduce the probability density
\[
 \rho_\theta(s)=e^{\theta s-\Lambda(\theta)}\rho(s).
\]
Differentiation gives
\[
 \Lambda'(\theta)=\int_{\mathbb R}s\rho_\theta(s)\,ds,
 \qquad
 \Lambda''(\theta)
 =\int_{\mathbb R}(s-\Lambda'(\theta))^2\rho_\theta(s)\,ds>0.
\]
The variance is positive because a probability distribution with a density
cannot be concentrated at a single point. In particular,~$\Lambda'$ is
strictly increasing and takes values in~$(\bar s_-,\bar s_+)$.

\medskip
\noindent\emph{The range of~$\Lambda'$.}
At a finite endpoint of~$D$, Fatou's lemma and the definition of~$D$ give
\begin{equation}\label{26oct2315}
\begin{aligned}
 &\lim_{\theta\to-a^+}\Lambda(\theta)=+\infty
       &&\text{if }a<+\infty,\\
 &\lim_{\theta\to b^-}\Lambda(\theta)=+\infty
       &&\text{if }b<+\infty.
\end{aligned}
\end{equation}
By convexity,~$\Lambda'$ tends to~$-\infty$ at a finite left endpoint
and to~$+\infty$ at a finite right endpoint.
Moreover, $a<+\infty$ forces $\bar s_-=-\infty$, since a distribution
bounded below has all negative exponential moments. Similarly,
$b<+\infty$ forces $\bar s_+=+\infty$.

Suppose now that $b=+\infty$. For every $c<\bar s_+$, we have
\[
 p_c:=\int_c^{+\infty}\rho(s)\,ds>0.
\]
For $\theta>0$, it follows that
\[
 \Lambda(\theta)\ge c\theta+\log p_c.
\]
Since $\Lambda(0)=0$, convexity yields
\[
 \Lambda'(\theta)\ge\frac{\Lambda(\theta)}{\theta}
 \ge c+\frac{\log p_c}{\theta}.
\]
Letting $\theta\to+\infty$ and then $c\to\bar s_+$ shows that
$\Lambda'(\theta)\to\bar s_+$, also when $\bar s_+=+\infty$.
Applying the same argument to the density of~$-s$ handles the case
$a=+\infty$. We have therefore proved
\[
 \lim_{\theta\to-a^+}\Lambda'(\theta)=\bar s_-\,,\qquad
 \lim_{\theta\to b^-}\Lambda'(\theta)=\bar s_+\,,
\]
with the natural interpretation at infinite endpoints.
Together with~$\Lambda''>0$, these limits show that~$\Lambda'$ is an
increasing $C^\infty$ diffeomorphism from~$D$ onto~$D'$.

\medskip
\noindent\emph{The rate function on~$D'$.}
For $m\in D'$, let $\bar\theta(m)$ be the unique solution of
$\Lambda'(\bar\theta(m))=m$.
Strict convexity of~$\Lambda$ shows that~$\bar\theta(m)$ is the unique
maximizer of~$\theta m-\Lambda(\theta)$ on~$D$.
Outside~$D$ this expression is~$-\infty$, so
\[
 I(m)=m\bar\theta(m)-\Lambda(\bar\theta(m)).
\]
The inverse function theorem shows that~$\bar\theta$ is $C^\infty$,
and hence so is~$I$ on~$D'$. Differentiating gives
\[
 I'(m)=\bar\theta(m)
   +\bigl(m-\Lambda'(\bar\theta(m))\bigr)\bar\theta'(m)
 =\bar\theta(m),
\]
and
\[
 I''(m)=\bar\theta'(m)
 =\frac{1}{\Lambda''(\bar\theta(m))}>0.
\]

\medskip
\noindent\emph{The rate function outside~$D'$.}
Suppose that $\ell:=\bar s_-$ is finite. Then $(-\infty,0]\subset D$,
and, for $\theta\le0$,
\[
 e^{\Lambda(\theta)-\ell\theta}
 =\int_\ell^{+\infty}e^{\theta(s-\ell)}\rho(s)\,ds\le1.
\]
Since the distribution has a density, it has no atom at~$\ell$.
Dominated convergence therefore gives
\[
 \Lambda(\theta)-\ell\theta\longrightarrow-\infty
 \qquad\text{as }\theta\to-\infty.
\]
For every $m\le\ell$,
\[
 \theta m-\Lambda(\theta)
 =\theta(m-\ell)-\bigl(\Lambda(\theta)-\ell\theta\bigr)
 \longrightarrow+\infty,
\]
so $I(m)=+\infty$. The same argument applied to~$-s$ shows that
$I(m)=+\infty$ for $m\ge\bar s_+$ whenever $\bar s_+$ is finite.
This proves~\eqref{26oct3026} and completes the proof.
\end{proof}

\end{document}